\documentclass[11pt,tightenlines,superscriptaddress,floatfixolumn,english,aps,notitlepage,nofootinbib]{revtex4-1}

\usepackage{amsmath}
\usepackage{esint}
\usepackage{graphicx}
\usepackage{hyperref}
\usepackage{color}
\usepackage{xcolor}
\usepackage[utf8]{inputenc} 
\advance\voffset -0.2in

\newcommand{\be}{\begin{eqnarray}}
\newcommand{\ee}{\end{eqnarray}}
\newcommand{\non}{\nonumber\\}

\newcommand{\ave}[1]{\left\langle #1 \right\rangle}

\newcommand{\gev}{{\rm \, GeV}}

\definecolor{vkcolor2}{HTML}{336600}

\begin{document}

\title{
Rapidity dependence of proton cumulants and correlation functions
}

\author{Adam Bzdak}
\email{bzdak@fis.agh.edu.pl}
\affiliation{AGH University of Science and Technology,
Faculty of Physics and Applied Computer Science,
30-059 Krak\'ow, Poland}

\author{Volker Koch}
\email{vkoch@lbl.gov}
\affiliation{Nuclear Science Division, 
Lawrence Berkeley National Laboratory, 
Berkeley, CA, 94720, USA}

\begin{abstract}
The dependence of multi-proton  correlation functions and
cumulants on the acceptance in rapidity and transverse momentum is studied. 
We find that the preliminary data of various cumulant ratios
are consistent, within errors, with rapidity and transverse momentum independent correlation
functions.
However, rapidity correlations which moderately increase with
rapidity separation between protons are slightly favored. We propose to further
explore the rapidity dependence of multi-particle correlation functions
by measuring the dependence of the integrated reduced correlation
functions as a function of the size of the rapidity window.    
\end{abstract}

\maketitle

\section{Introduction}
\label{Sec:Introduction}

One of the central goals in strong interaction research is to explore
the phase diagram of QCD. Of particular interest is the search for
a possible first-order phase coexistence region and its
associated critical point. A significant effort in this search, experimentally as well as
theoretically, is concentrating on the measurements and calculations
of correlations and cumulants of conserved charges. A particular
emphasis has been put on the cumulants of the baryon number 
 \cite{Stephanov:1998dy,Stephanov:2008qz,Skokov:2010uh,Friman:2011pf,Adamczyk:2013dal,Adamczyk:2014fia}, see also
 \cite{Borsanyi:2011sw,Bazavov:2012jq,Bellwied:2015lba,
 Gavai:2010zn,Kitazawa:2013bta,Asakawa:2015ybt,Mukherjee:2016nhb,Herold:2016uvv,
 Chatterjee:2016mve,Lacey:2016tsw,Hippert:2017xoj,Rougemont:2017tlu,Almasi:2017bhq,He:2017zpg}
(see, e.g., \cite{Koch:2008ia} for an overview). Interpreting these
higher order cumulants and their measurements, however, is not a
straightforward exercise as discussed, e.g., in \cite{Bzdak:2012an,Skokov:2012ds,Bzdak:2012ab,Bzdak:2013pha,Luo:2014rea,Nonaka:2016xje,Westfall:2014fwa,
Feckova:2015qza,Bzdak:2016qdc,Braun-Munzinger:2016yjz,Bluhm:2016byc,Xu:2016skm,Nonaka:2017kko,Garg:2017agr}.
Also, different, although related, ideas, based on an intermittency analysis in the
transverse momentum phase space have been explored
\cite{Anticic:2009pe,Anticic:2012xb,Antoniou:2015lwa}.

Recently, it has been pointed out \cite{Ling:2015yau,Bzdak:2016sxg}, 
(see also \cite{Kitazawa:2017ljq,He:2017zpg}), that it may be
more instructive to study (integrated) multi-particle correlations
instead of cumulants. In the limit when anti-particles can be ignored,
which is the case for anti-protons at low beam energies, the integrated multi-particle
correlations are linear combinations of the various cumulants and
thus can be extracted easily from the measured cumulants. This has
been done on the basis of preliminary data on proton cumulants from the STAR collaboration
\cite{Luo:2015ewa}. It was found that the systems created at low beam
energies ($7.7 - 11.5 \gev$) exhibit sizable three- and strong four-proton correlations
\cite{Bzdak:2016sxg,Luo:2017faz}. Indeed, as pointed out in
\cite{Bzdak:2016jxo}, in order to reproduce the observed magnitude of
these correlations one has, for example, to assume 
a strong presence of eight-nucleon (or four-proton) clusters in the
system. In addition to the sheer magnitude of the correlations, the
centrality and rapidity dependence of these correlations give additional
insight into properties of the systems created in these collisions \cite{Bzdak:2016sxg}. 

In this paper we will explore the rapidity and to some extent transverse momentum
dependence of multi-particle correlations in more detail. One of our motivations is a recent
preliminary observation by the STAR
collaboration \cite{star:a1,star:a2} regarding the rapidity dependence of the two-proton
correlation function. Within the rapidity window $|y|<0.8$, STAR finds that  
across all RHIC energies the two proton reduced correlation function (see the definition in
section \ref{Sec:Notation}) in central Au+Au collisions is strongly increasing with the rapidity
separation, $y_{1}-y_{2}$, between the two protons. The
shape of the correlation function can be approximately described by 
\begin{equation}
c_{2}(y_{1}-y_{2})=c_{2}^0 + \gamma_{2}\left( y_{1}-y_{2}\right) ^{2},\qquad \gamma_{2}>0, 
\label{eq:c2-star}
\end{equation}%
where $c_{2}^0$ is the value at $y_{1}-y_{2}=0$, and $\gamma_{2}$ is a positive
number, with $\gamma_{2}\sim 2\times 10^{-2}$ at $\sqrt{s}=7.7 \gev$
\cite{star:a1}.\footnote{At the recent CPOD conference STAR reported \cite{star:cpod}
  that the rapidity dependence of the two-proton correlation
  function depends considerably on the method employed to subtract the
  uncorrelated single particle contribution from the data. Thus the
  value for $\gamma_{2}$ quoted here may still change and should
  be taken only as a rough guidance.} Taking such a correlation 
at face value, one would conclude that protons prefer to be separated in rapidity, or, 
in other words, they
seem to repel each other. The shape of the correlation function
is roughly energy independent, which is 
rather surprising since protons at, say, $7.7$ GeV, originate almost
exclusively from
the target and projectile nuclei whereas at $200 \gev$, the protons 
at mid-rapidity mostly are produced.

The apparent anti-correlation between two protons was first observed in $e^{+}e^{-}$%
collisions at $\sqrt{s}=29$ GeV \cite{Aihara:1986fy}. Recently an analogous
observation was made by the ALICE collaboration in the context of the
two-baryon azimuthal correlations \cite{Adam:2016iwf}. This
measurement also found similar anti-correlations between protons and
lambdas, suggesting that the observed effects are not due to the Pauli
exclusion principle or electromagnetic interactions.
To our knowledge, the origin of this effect remains an open question,
which is important to resolve. Formation of clusters, as suggested in
\cite{Bzdak:2016jxo}, and as expected close to a critical point and
a phase transition, would naively lead to attractive
correlations in rapidity (i.e., protons would prefer to have similar rapidity) 
and not anti-correlations. However, we should keep in mind that these
correlations are in rapidity and not in configuration space. 
Also, one should  note that this effect, which, so far, is only  observed 
for two-particle correlations, may not be inconsistent with the
negative value for the integrated two-particle 
correlations extracted from the cumulant measurements
\cite{Bzdak:2016sxg, Luo:2017faz}.
In general, the  sign of an integrated multi-particle correlation also
  is driven by a pedestal. For example, in the case of two protons,
  $c_2^0$ in Eq. (\ref{eq:c2-star}) may depend on the fluctuations of the volume, or
  rather the number of wounded nucleons \cite{Bialas:1976ed,Skokov:2012ds,Bzdak:2016jxo,
  Braun-Munzinger:2016yjz}, and is not necessarily related to a possible repulsion or attraction 
  in rapidity between protons.
 
Clearly the rapidity dependence of the proton correlations need
  to be studied to gain further insight into the aforementioned sizable
  three-proton and strong four-proton correlations observed at low energies. It
  is the purpose of this paper to start exploring this issue. To this
  end we study the dependence of the multi-proton correlation
functions on rapidity, and, to some extent, on the  transverse
momentum. We show that the preliminary STAR data \cite{Luo:2015ewa} 
are consistent with constant multi-proton correlation functions and
slightly favor multi-proton anti-correlations in rapidity. We also demonstrate
that these correlations can be constrained further 
by measuring integrated reduced or normalized correlation functions as a function of the rapidity
window $\Delta y$.

This paper is organized as follows. In the next section we introduce
the notation and discuss the behavior of cumulants and correlation
functions in the limits of small and large acceptances. Next we analyze
the preliminary STAR data and extract some trends about the rapidity
dependence of three-proton and four-proton correlations. We also will
propose a means to extract more detailed information about the
multi-particle correlations. In the last section we conclude with a discussion of the
essential results.

\section{Notation and comments}
\label{Sec:Notation}

In this paper we focus on protons only and in the following we denote
the proton number by $N$ and its deviation from the mean by $\delta N
= N-\ave{N}$. Here $\ave{N}$ is the mean
number of protons at a given centrality. The cumulants of the proton
distribution function as measured by the STAR collaboration are
then given by 
\begin{equation}
K_{1}\equiv \langle N\rangle ;\quad K_{2}\equiv \langle (\delta
N)^{2}\rangle ;\quad K_{3}\equiv \langle \left( \delta N\right) ^{3}\rangle
;\quad K_{4}\equiv \langle \left( \delta N\right) ^{4}\rangle -3\langle
(\delta N)^{2}\rangle ^{2}.  \label{eq:K1234}
\end{equation}

As already eluded to in the Introduction, the cumulants can be expressed in terms of
the multi-particle integrated correlation functions
\cite{Bzdak:2016sxg}, which also are known as factorial cumulants \cite{Ling:2015yau}
\begin{eqnarray}
K_{2} &=&\left\langle N\right\rangle +C_{2},  \label{eq:K2C} \\
K_{3} &=&\left\langle N\right\rangle +3C_{2}+C_{3},  \label{eq:K3C} \\
K_{4} &=&\left\langle N\right\rangle +7C_{2}+6C_{3}+C_{4},  \label{eq:K4C}
\end{eqnarray}%
where
\begin{eqnarray}
C_{2} &=&\int dy_{1}dy_{2}C_{2}(y_{1},y_{2})  \notag \\
&=&\int dy_{1}dy_{2}\left[ \rho _{2}(y_{1},y_{2})-\rho (y_{1})\rho
(y_{2})\right] ,  \label{eq:C2}
\end{eqnarray}%
and similar for higher order correlation functions. See, e.g., Ref. \cite{Bzdak:2015dja}
for explicit definitions of the correlation functions up to the sixth order. In
Eq.~(\ref{eq:C2}) $C_{2}(y_{1},y_{2})$ is the two-particle rapidity
correlation function, $\rho _{2}(y_{1},y_{2})$ is the two-particle rapidity
density, and $\rho (y)$ is the single-particle rapidity distribution. The
generalization of Eqs.~(\ref{eq:K2C}-\ref{eq:K4C}) to two species of
particles can be found in the Appendix of Ref. \cite{Bzdak:2016sxg}. Here and in the
following $y_{i}$ denotes rapidity or in general, a set of variables under
consideration $(y_{i},p_{t,i},\varphi _{i})$.

It is a convenient and common practice to define the reduced correlation function%
\begin{equation}
c_{n}\left( y_{1},...,y_{n}\right) =\frac{C_{n}\left( y_{1},...,y_{n}\right) 
}{\rho \left( y_{1}\right) \cdots \rho \left( y_{n}\right) }.
\label{eq:cn-y}
\end{equation}%
The integral of the reduced correlation function over some given
acceptance range, we subsequently will call, for the lack of
a better term, ``coupling'' %
\begin{equation}
c_{n}=\frac{C_{n}}{\left\langle N\right\rangle ^{n}}=\frac{\int \rho \left(
y_{1}\right) \cdots \rho \left( y_{n}\right) c_{n}\left(
y_{1},...,y_{n}\right) dy_{1}\cdots dy_{n}}{\int \rho \left( y_{1}\right)
\cdots \rho \left( y_{n}\right) dy_{1}\cdots dy_{n}}.  \label{eq:coupling}
\end{equation}%
The cumulants $K_{n}$ then may be expressed in terms of the
couplings $c_{n}$,
\begin{eqnarray}
K_{2} &=&\left\langle N\right\rangle +\left\langle N\right\rangle ^{2}c_{2},
\label{eq:K2-cn} \\
K_{3} &=&\left\langle N\right\rangle +3\left\langle N\right\rangle
^{2}c_{2}+\left\langle N\right\rangle ^{3}c_{3},  \label{eq:K3-cn} \\
K_{4} &=&\left\langle N\right\rangle +7\left\langle N\right\rangle
^{2}c_{2}+6\left\langle N\right\rangle ^{3}c_{3}+\left\langle N\right\rangle
^{4}c_{4}.  \label{eq:K4-cn}
\end{eqnarray}

Of course, mathematically, the cumulants $K_{1}=\langle
N\rangle$, $K_2$, $K_3$, and $K_4$ carry exactly 
the same information as [$C_2$, $C_{3}$, $C_4$] or [$c_2$, $c_3$,
$c_4$]. 
However, as already discussed in \cite{Bzdak:2016sxg}, studying
cumulants may not be the best way to extract information about the
dynamics of the system, since: 
(i)  cumulants mix the correlation functions of different orders and 
(ii) they might be dominated by a trivial term $\langle N\rangle$ even in the presence 
of interesting dynamics. 

One such example, where the trivial term  $\langle N\rangle$ dominates
and thus hides the interesting physics is the limit of small
acceptance as we will discuss next.

\subsection{Effective Poisson limit}
\label{sec:poisson}
Before we discuss the rapidity and transverse momentum dependence of
the various cumulants and correlations, let us briefly remind ourselves
what happens if one considers the limit of small or vanishing
acceptance. Here, we will restrict ourselves to correlations in rapidity,
however our arguments will be general and apply to any variables. 
Suppose that particles are measured in a rapidity interval
$y_{0}\leq y \leq y_{0}+ \Delta y$ and that $%
\Delta y\rightarrow 0$. 
Let us first consider two-particle correlations. For sufficiently small $%
\Delta y$ any reasonable correlation function $c_{2}(y_{1},y_{2})$ may be approximated by a
constant.\footnote{%
For the extreme case of $c_{2}(y_{1},y_{2})\sim\delta (y_{1}-y_{2})$, $c_{2}$, given 
by Eq.~(\ref{eq:coupling}), depends on the
acceptance window even for very small rapidity intervals, and our argument does not apply.
However, a Dirac delta correlation function is of no interest in any
practical situation.} As a consequence, for sufficiently small $\Delta y$, the coupling,  $c_{2}$, is
independent of  $\Delta y$, as can be seen from Eq.~(\ref{eq:coupling}). In
other words, suppose that $c_{2}(y_{1},y_{2})%
\simeq c_{2}^{0}$ for very small $\Delta y$, then 
\begin{equation}
c_{2}=\frac{\int_{\Delta y}\rho (y_{1})\rho(y_{2})c_{2}(y_{1},y_{2})
dy_{1}dy_{2}}{\int_{\Delta y}\rho (y_{1})\rho(y_{2})
dy_{1}dy_{2}}\simeq c_{2}^{0}.
\end{equation}
We emphasize that $c_{2}^{0}$ may assume any value. 
However, whatever the value of $c_{2}^{0}$, in the limit of 
$\Delta y\rightarrow 0$, we have $\left\langle N\right\rangle \rightarrow
0$ and $K_{2}\simeq \left\langle N\right\rangle $ (see Eq.~\eqref{eq:K2-cn}). Exactly the
same argument holds for any $K_{n}$, and we obtain $K_{n}\simeq \left\langle
N\right\rangle $ and consequently all cumulant ratios equal to unity,
$K_{n}/K_{m}\simeq 1$. 

Therefore, even in the presence of  sizable correlations,
their effects on the cumulants are suppressed for small
acceptance. 
Actually, as can be seen from
Eqs.~\eqref{eq:K2-cn}-\eqref{eq:K4-cn}, it is the number of particles
which determines if the cumulants are dominated by $\langle N\rangle$
and, thus, their ratios are close to unity. 
For example, if $\langle N\rangle ^4 c_4 <<
    \langle N\rangle$, the fourth order cumulant, $K_4$, is
    practically not sensitive to four-proton correlations even if
    $c_4$ is different from zero and may carry some interesting information. 
    Therefore, even for large acceptance the cumulants
    are close to the Poisson limit if one is dealing with rare
    particles. This may very well be the reason that for low energies
    STAR observes a cumulant ratio of $K_{4}/K_{2}\simeq 1$ for
    anti-protons, and it would be interesting to measure the couplings, $c_n$,
    for anti-protons in order to see if  anti-protons exhibit the same correlations as
    protons at low energies. 
 
Clearly measuring cumulants and looking for the deviation from the
Poisson limit is not the most optimal way to extract possible non-trivial 
correlations resulting from criticality etc. Instead, one either should directly measure the
differential multi-particle correlation (Eq.~\eqref{eq:cn-y}) or, at
the very least, extract the couplings, $c_{n}$,
Eq.~\eqref{eq:coupling}. Their 
dependence on the acceptance does reflect a change in physics and is
not simply a consequence of a change in the number of 
particles.\footnote{An additional 
  advantage of the couplings is that they
  are independent of the efficiency of the detector as long as the
  efficiency follows a binomial distribution and is phase space independent
  \cite{Bzdak:2012ab,Bzdak:2016qdc,Kitazawa:2017ljq}.}

After having investigated the case of small acceptance let us next
turn to the opposite limit of (nearly) full acceptance.

\subsection{Full acceptance}

Let us next study what happens in the situation when all baryons,
including the spectators, are detected. 
In this case (again, we
consider low energies and neglect anti-baryons)  $N=\langle
N\rangle=B$, where $B$ is the total baryon number of the entire system. Therefore, $\delta N=0$ and obviously $K_n=0$ for $n \ge 2$. Using Eqs. (\ref{eq:K2C}-\ref{eq:K4C}) and (\ref{eq:K2-cn}-\ref{eq:K4-cn}) we obtain
\begin{equation}
C_{2}=-B,\qquad C_{3}=2B,\qquad C_{4}=-6B,  
\label{eq:C-max}
\end{equation}
and
\begin{equation}
c_{2}=-\frac{1}{B},\qquad c_{3}=\frac{2}{B^{2}},\qquad c_{4}=-\frac{6}{B^{3}}. 
\label{eq:c-max}
\end{equation}
We note that this is a general result and it is insensitive to the
presence of any dynamics other than global baryon number
conservation. 

Finally let us note that $K_3/K_2 \rightarrow -1$ and $K_4/K_2
\rightarrow 1$  when we approach the limit of full acceptance. To see
this let us consider a region in phase space, denoted by $(a)$,
and the remaining phase space, or complement, which we denote by
$(b)$. Since the baryon number is conserved, having 
$N_{(a)}$ baryons in region $(a)$ implies $N_{(b)}=B-N_{(a)}$ baryons
in the complement, $(b)$. Since $\delta B = 0$ we have 
\begin{equation}
\delta N_{(b)} = \delta(B-N_{(a)}) = -\delta N_{(a)} , 
\end{equation}
and consequently 
\begin{eqnarray} 
K_{n,(a)} &=& K_{n,(b)} \quad n=2,4,6,..., \notag \\
K_{n,(a)} &=& -K_{n,(b)} \quad n=3,5,7,....
\label{eq:relation}
\end{eqnarray}
Here $K_{n,(a)}$ is the cumulant measured 
in region $(a)$ and $K_{n,(b)}$ is the cumulant in a remaining
part of the full phase space, $(b)$. This is a rather nontrivial and general
consequence of baryon conservation. A more rigorous derivation is
presented in the Appendix.

In the previous subsection we argued that for very small acceptance the
cumulant ratio goes to $1$ and thus the cumulant ratio for the full
acceptance goes to $-1$ for $K_{3}/K_{2}$ and to $1$ for
$K_{4}/K_{2}$. 
The integrated correlation functions and the couplings, on the other
hand, do not show such a symmetry between a given region of phase space
and its compliment.
This is shown in detail in the Appendix but can
already be inferred from the fact that in the limit of full
acceptance the couplings are determined entirely by the total
baryon number $B$. In the limit of vanishing acceptance, however, other
physics also affects the value of the couplings, as discussed in
Section~\ref{sec:poisson}.

Having discussed the limits of small and full acceptances we now turn
to the rapidity dependence of the cumulants and correlation functions.

\section{Results}
\label{Sec:Results}

In this section we discuss in detail the rapidity and, to some extent, the transverse
momentum dependence of multi-proton cumulants and correlation
functions. First, we will explore the limit of rapidity and transverse
momentum independent correlations. Next we will discuss to which
extent the present preliminary STAR data allow us to set limits on the
rapidity dependence of the underlying correlations.

\subsection{Constant correlation}

\begin{figure}[t]
\begin{center}
\includegraphics[scale=0.35]{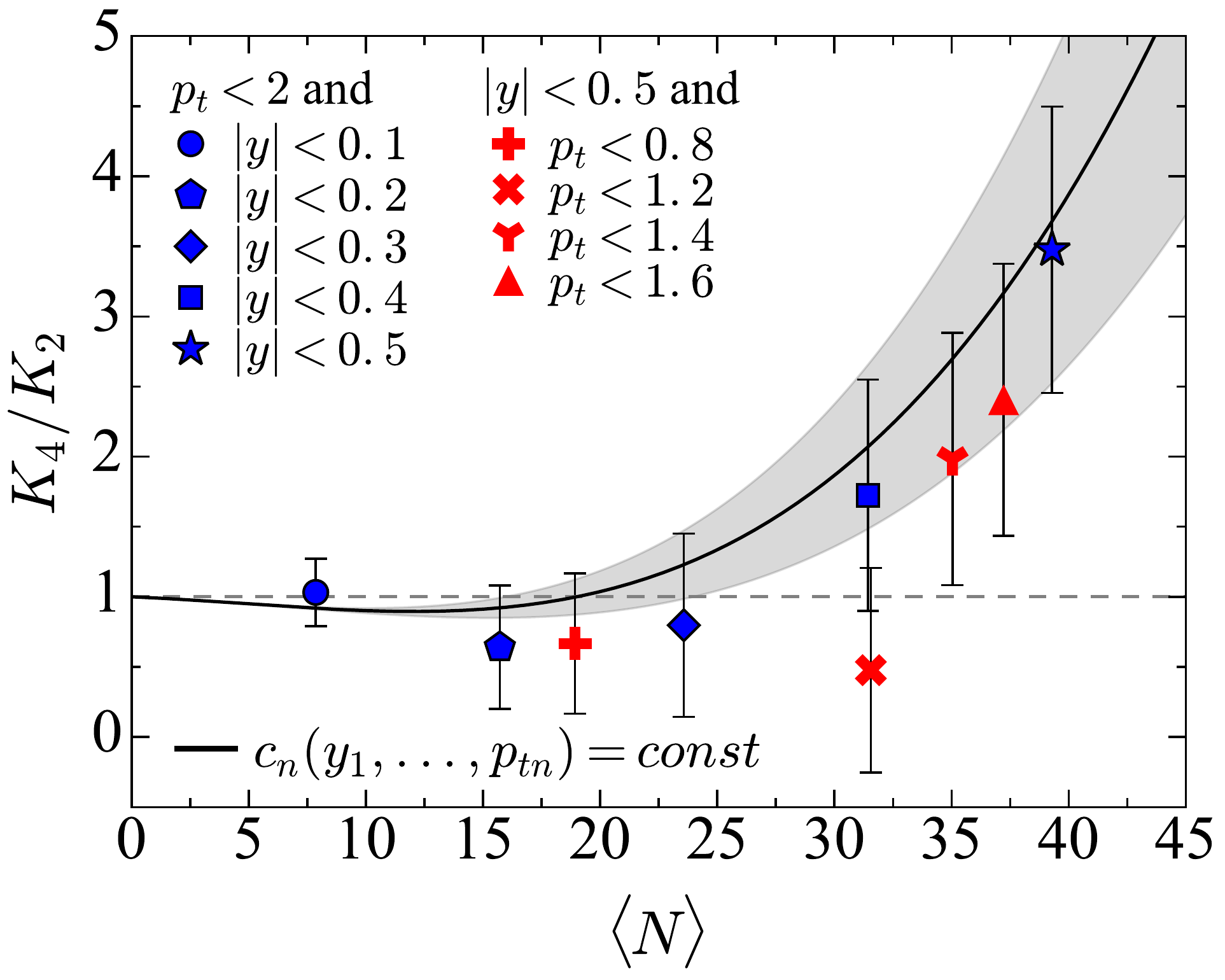}
\end{center}
\par
\vspace{-5mm}
\caption{The cumulant ratio $K_{4}/K_{2}$ in central $0-5\%$ Au+Au
  collisions at $\sqrt{s}=7.7$ GeV as a function of the number of
  measured protons, $\left\langle N\right\rangle $, for different
  acceptance windows in rapidity and transverse momentum (in units of GeV). 
  For all data points $p_{t}>0.4 \gev$. The black solid line represents a prediction 
  based on a constant correlation function, see Eq. (\ref{eq:cn-const}). 
  The shaded band is driven mostly by the large experimental uncertainty of $K_4$. 
  Based on preliminary STAR data \protect\cite{Luo:2015ewa}.}
\label{fig:k4k2_scaling}
\end{figure}
Let us start with the simplest assumption namely that  the reduced
correlation function does not depend on rapidity and transverse momentum, i.e., 
\begin{equation}
c_{n}(y_{1},p_{t1},...,y_{n},p_{tn})=const=c_{n}^{0}.  \label{eq:cn-const}
\end{equation}
This rather extreme assumption, however, is, as we will show below,  
consistent with the preliminary STAR data at $7.7$ GeV (see also
\cite{Bzdak:2016sxg}). In addition, in this case the couplings
$c_{n}$ do not depend  on rapidity and transverse momentum either,
as can be seen from Eq. \eqref{eq:coupling}
\begin{equation}
c_{n}=c_{n}^{0}.
\end{equation}%
The multi-particle integrated correlation functions, $C_{n}=\langle N\rangle^n c_n$, 
and cumulants, $K_{n}$, in turn depend
on the acceptance only through their dependence on the number of
protons $\left\langle N\right\rangle $, see Eqs. (\ref{eq:K2-cn}-\ref{eq:K4-cn}).
Therefore, in  Fig. \ref{fig:k4k2_scaling} we plot $K_{4}/K_{2}$ as measured by STAR as
a function of $\left\langle N\right\rangle $ for different rapidity and
transverse momentum intervals. 

The black solid line in Fig. \ref{fig:k4k2_scaling} represents a prediction  based on a constant correlation
function.  In this calculation we have three unknown parameters, $c_{2}^{0},$ 
$c_{3}^{0}$ and $c_{4}^{0}$. Since these numbers do not depend on acceptance
we determine them from the preliminary data for $|y|<0.5$ ($\Delta y=1$) and 
$0.4<p_{t}<2$ GeV, that is, from the maximal acceptance currently
available.
Here we use Eqs. (\ref{eq:K2-cn}-\ref{eq:K4-cn}) and the values for 
$\langle N\rangle$, $K_{2}$, $K_{3}$ and $K_{4}$ shown in
Ref. \cite{Luo:2015ewa}.\footnote{We determine $c_n^0$ from the proton
  cumulants but compare to $y$ and $p_t$ dependence of the net-proton
  cumulants, which are the only data currently available. Although at $7.7$
  GeV the number of anti-protons is practically negligible, it results in 
  a slight disagreement of the black solid line with the blue star 
  in Fig. \ref{fig:k4k2_scaling}.}
To determine $\left\langle N\right\rangle $ at a given acceptance region
we assume the single proton rapidity distribution to be flat as a function of
rapidity, i.e., $\left\langle N\right\rangle =\left\langle N_{\Delta
y=1}\right\rangle \Delta y$ and for the transverse momentum single proton
distribution we take $\rho (p_{t})\sim p_{t}\exp (-m_{t}/T)$ with $T=0.27$
GeV and $m_{t}=(m^{2}+p_{t}^{2})^{1/2}$ with $m=0.94$ GeV. Both these
assumptions are well supported by experimental data
\cite{Adamczyk:2017iwn,Anticic:2010mp}.
Having $c_{n}^{0}$, we can predict the cumulants or the correlation functions for any
acceptance characterized by $\left\langle N\right\rangle $, whether in
transverse momentum or in rapidity.\footnote{Based on the preliminary
  STAR data for the cumulants \cite{Luo:2015ewa} we obtain 
  $c_2^0 \approx -1.1\times 10^{-3}$, 
  $c_3^0 \approx -1.7\times 10^{-4}$ and $c_4^0 \approx 7.3\times 10^{-5}$.} 

Interestingly we find that, except for one point at $|y|<0.5$ and $0.4<p_{t}<1.2$ GeV, all the
points follow, within the admittedly large experimental error bars, one universal curve
consistent with a constant correlation function. The fact that the
rapidity dependence of the cumulant ratio $K_{4}/K_{2}$ is consistent
with long-range rapidity correlations already has been found in
\cite{Bzdak:2016sxg}. That the transverse momentum dependence is also
consistent with long-range correlations is new. If correct, it would,
for example, imply that the cumulant ratio $K_{4}/K_{2}$ has
roughly the
same value (close to unity) for a transverse momentum range of
$0.8 \gev<p_{t}<2 \gev$ as the value for the range of $0.4 \gev < p_{t}
<0.8\gev$ since, in both $p_t$ windows, $\langle N\rangle$ is
approximately the same. The result for the $p_{t}$-range of $0.4 \gev < p_{t}
<0.8\gev$ has been published by the STAR collaboration in \cite{Adamczyk:2013dal}. 

Of course, the error bars in the preliminary STAR data are rather
sizable and, therefore, a mild dependence of the correlation function
on rapidity (and transverse momentum) cannot be ruled out. In
addition, as already mentioned in the Introduction, the preliminary,
explicit measurement of the two-proton correlation function
\cite{star:a1,star:a2} does exhibit an increase with increasing rapidity 
difference of a proton pair, $y_1 - y_2$. To explore this further we next will 
allow for some mild rapidity dependence of the correlation function.

\subsection{Rapidity dependent correlation}

\begin{figure}[t]
\begin{center}
\includegraphics[scale=0.32]{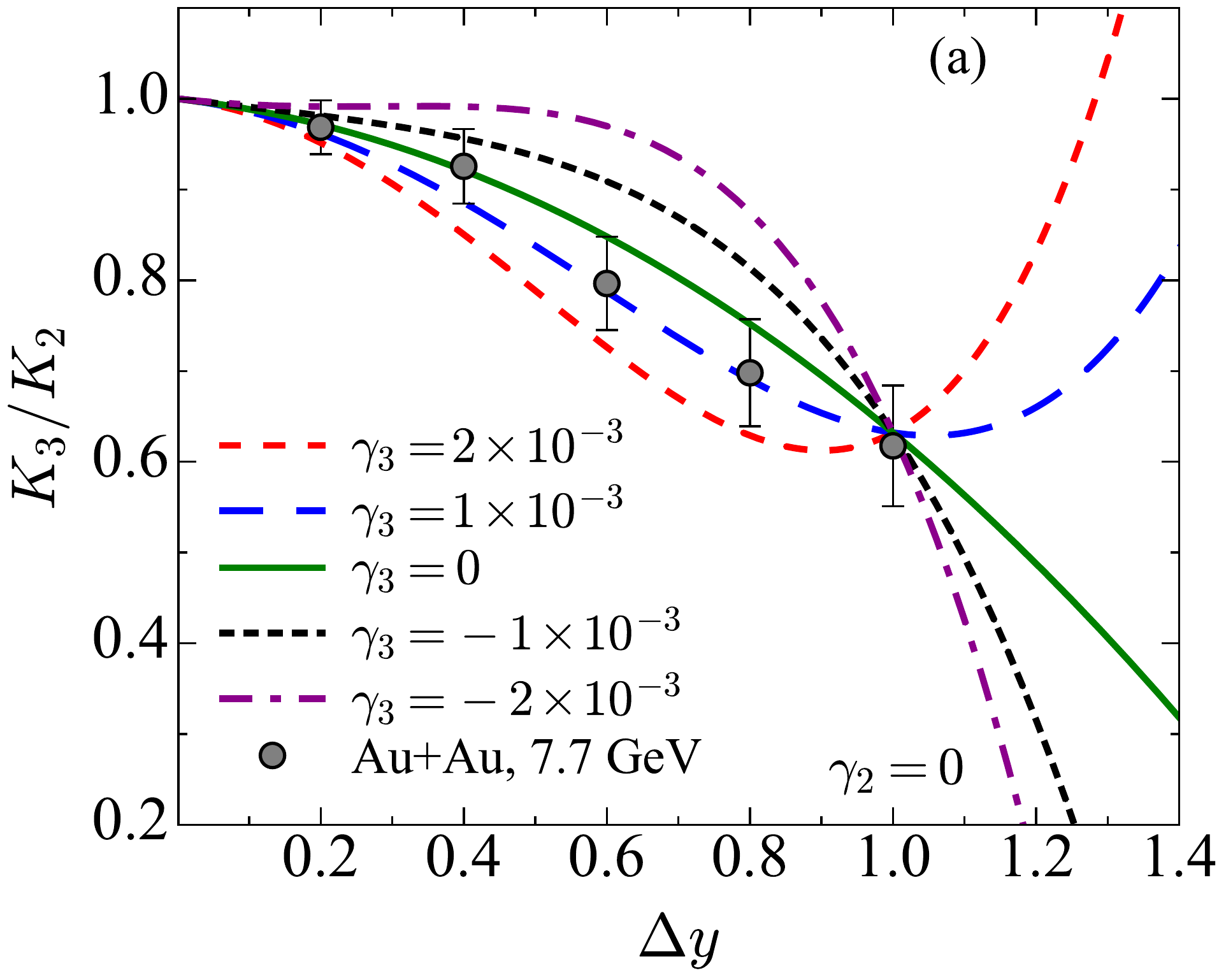} %
\includegraphics[scale=0.32]{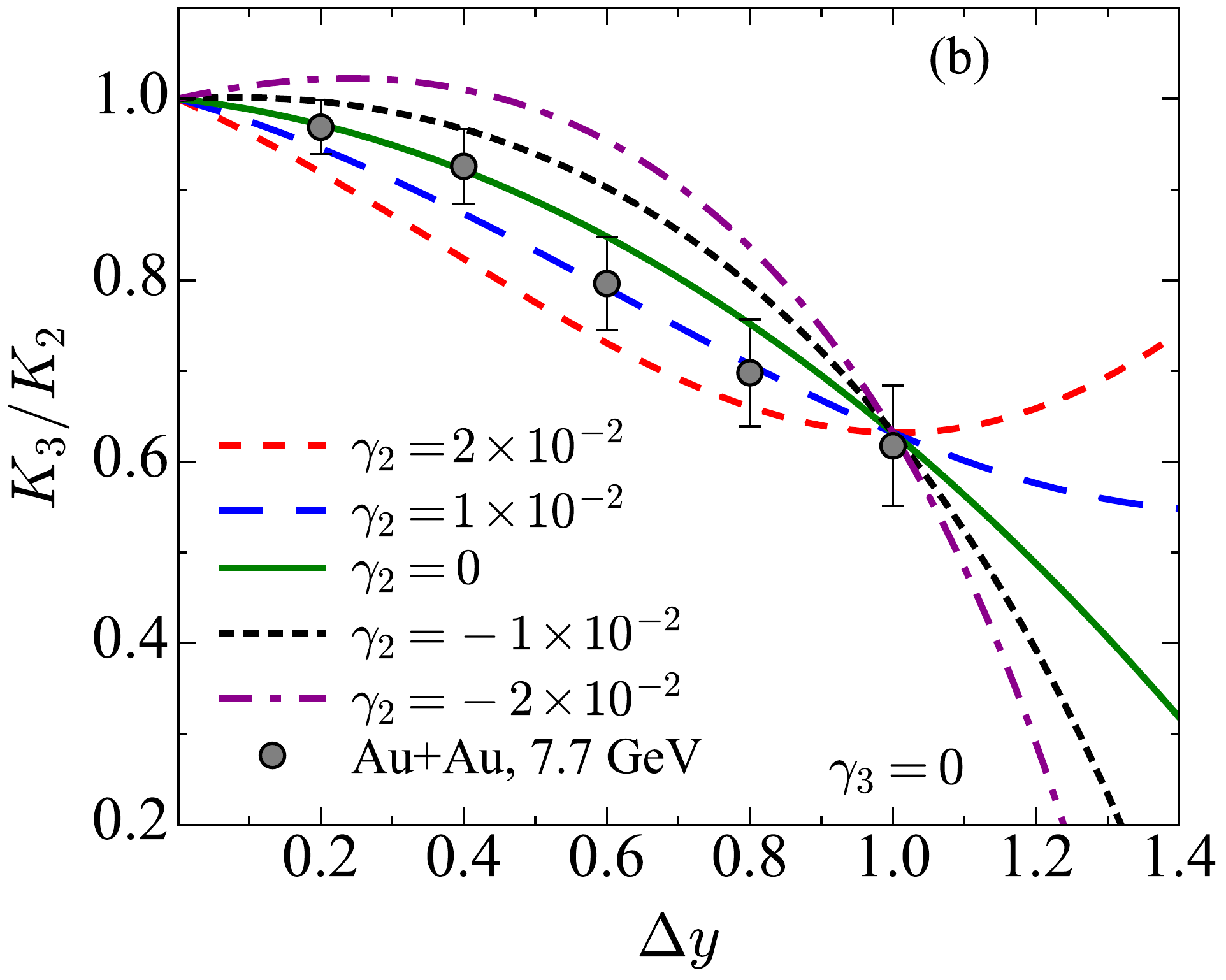}
\end{center}
\par
\vspace{-5mm}
\caption{The cumulant ratio $K_{3}/K_{2}$ in central Au+Au collisions
  at $\sqrt{s}=7.7$ GeV as a function of the rapidity acceptance $\Delta y$, $|y|<\Delta y/2$, 
  for (a) $\protect\gamma _{2}=0$ and different values of
  $\protect\gamma _{3}$ from Eq. (\ref{eq:c234-gamma}) and (b)
  $\protect\gamma _{3}=0$ and different values of $\protect\gamma
  _{2}$.  
  Based on preliminary STAR data \protect\cite{Luo:2015ewa}.}
\label{fig:k3k2_c23}
\end{figure}

In the previous subsection we demonstrated that the STAR data for $%
K_{4}/K_{2}$ at $7.7$ GeV are consistent with a constant multi-proton
correlation function. Here we study how sensitive the cumulant ratios
and correlations are to a certain (weak) rapidity dependence. To this
end we consider the leading correction to a constant correlation
function, which should be even in $y_i-y_k$. Thus we explore the
following ans\"atze for the reduced correlation functions  
\begin{eqnarray}
c_{2}(y_{1},y_{2}) &=&c_{2}^{0}+\gamma _{2}\left( y_{1}-y_{2}\right) ^{2}, 
\notag \\
c_{3}(y_{1},y_{2},y_{3}) &=&c_{3}^{0}+\gamma _{3}\tfrac{1}{3}\left[ \left(
y_{1}-y_{2}\right) ^{2}+\left( y_{1}-y_{3}\right) ^{2}+\left(
y_{2}-y_{3}\right) ^{2}\right] ,  \notag \\
c_{4}(y_{1},y_{2},y_{3},y_{4}) &=&c_{4}^{0}+\gamma _{4}\tfrac{1}{6}\left[
\left( y_{1}-y_{2}\right) ^{2}+\left( y_{1}-y_{3}\right) ^{2}+\left(
y_{1}-y_{4}\right) ^{2}\right.   \notag \\
&&\,\,\,\,\,\,\,\,\,\,\,\,\,\,\,\,\,\,\,\,\,\,\,\,\,\left. +\left( y_{2}-y_{3}\right) ^{2}+\left( y_{2}-y_{4}\right)
^{2}+\left( y_{3}-y_{4}\right) ^{2}\right] ,
\label{eq:c234-gamma}
\end{eqnarray}%
where $\gamma _{n}$ measures the deviation from $%
c_{n}(y_{1},...,y_{n})=const $. Note that we have constructed the
correlation function such that positive values of $\gamma _{n}$ result in
growing correlations with rapidity separation between particles. We
further note that the above form for the two-proton reduced correlation function, 
$c_{2}(y_{1},y_{2})$, is supported by the preliminary STAR data \cite{star:a1,star:a2} 
where $\gamma _{2}>0$, that is, two protons do not want to occupy
  the same rapidity. Our simple formulas for $c_3$ and $c_4$ are not
  supported by any known data, however, we believe they should serve as
  a reasonable representation for the correlation if the distance in rapidity
  between protons is not too large. 
Within the region of validity of our simple ansatz, the coefficients $\gamma _{n}$
  have a clear physical interpretation, and here we will 
  constrain their values or at least their signs. To this end we will
  use the preliminary STAR data for
  $K_{3}/K_{2}$ and $K_4/K_2$. Although, as already pointed out, the
  rapidity dependence of these cumulant ratios is consistent with
  constant correlations, we will see that the data allow for excluding certain values for
  $\gamma_{n}$ and possibly even determine their sign. 

Taking the above relations and integrating in Eq. (\ref{eq:coupling})
over $|y_{i}|<\Delta y/2$ we obtain for the couplings 
\begin{equation}
c_{n}(\Delta y)=\frac{C_{n}}{\left\langle N\right\rangle ^{n}}=c_{n}^{0}+\gamma _{n}%
\frac{1}{6}(\Delta y)^{2}.  \label{eq:cn-delta-y}
\end{equation}
The couplings, $c_{n}(\Delta y)$, which depend on the region of acceptance, $\Delta y$ ($|y_{i}|<\Delta y/2$),
should not be confused with the reduced correlation functions, $c_n(y_1,...,y_n)$, which
depend on the rapidities of the individual particles.
As before, for a given  $\gamma _{n}$ the constant term,  $c_{n}^{0}$  is extracted from the STAR
data at $\Delta y=1$ ($|y|<0.5$) and $0.4<p_{t}<2$ GeV. Consequently, $c_{n}^{0}$ will depend on the
choice of $\gamma_{n}$.

\begin{figure}[t]
\begin{center}
\includegraphics[scale=0.32]{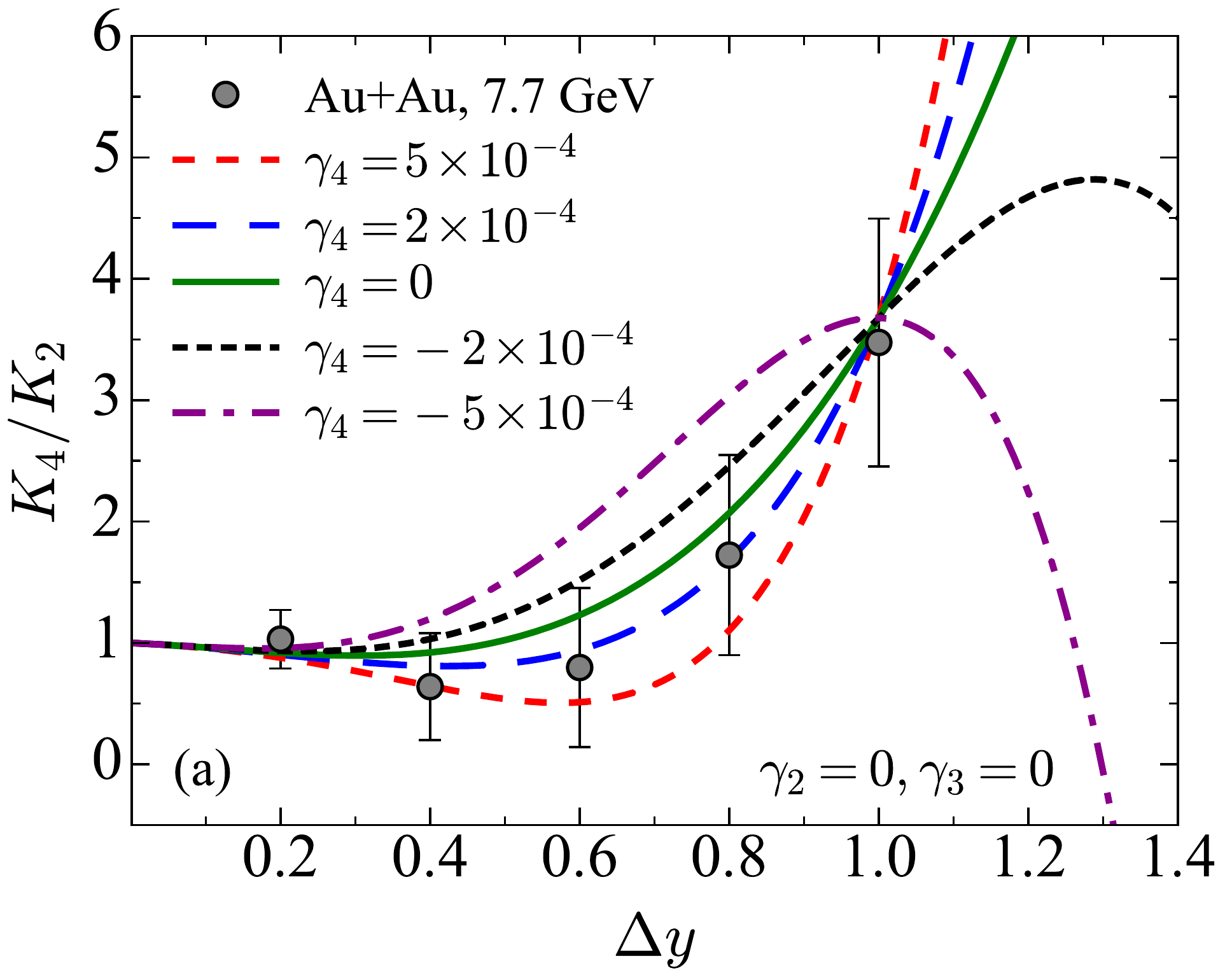} %
\includegraphics[scale=0.32]{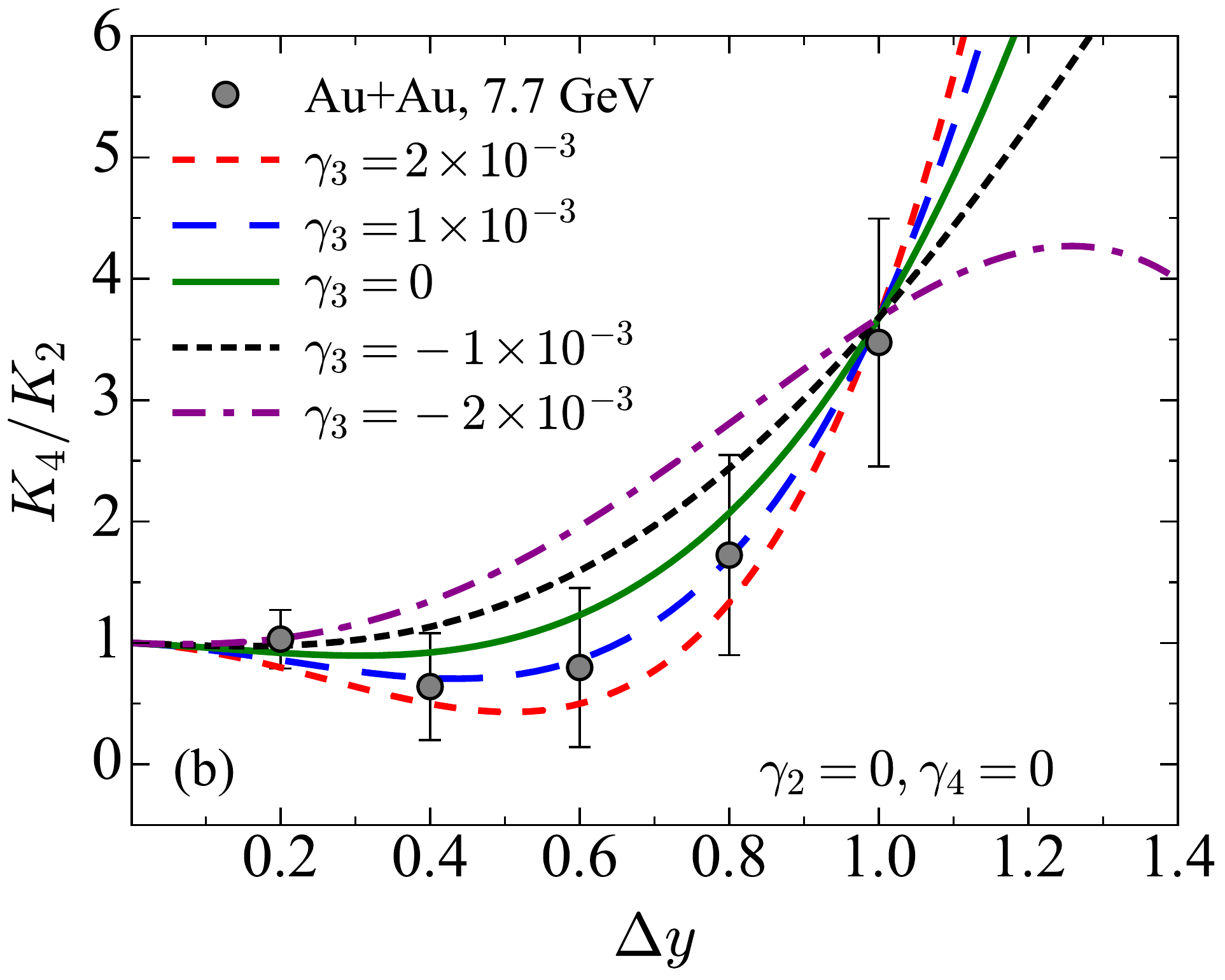} %
\includegraphics[scale=0.32]{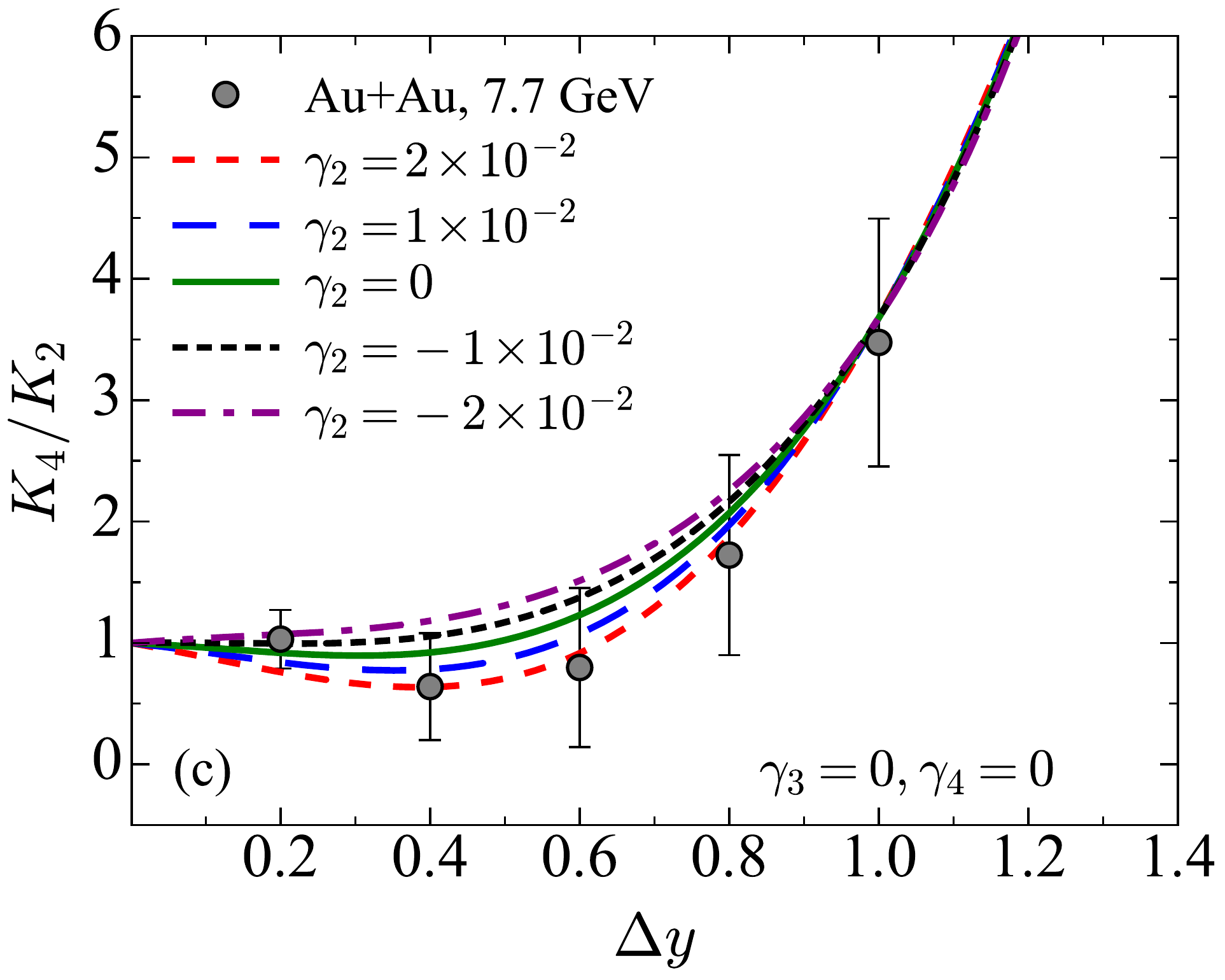}
\end{center}
\par
\vspace{-5mm}
\caption{The cumulant ratio $K_{4}/K_{2}$ in central Au+Au collisions
  at $\sqrt{s}=7.7$ GeV as a function of the rapidity acceptance $\Delta y$, $|y|<\Delta y/2$, for 
  (a) different values of $\protect\gamma _{4}$, 
  (b) $\protect\gamma _{3}$ and (c) $\protect\gamma _{2}$ from
  Eq. (\ref{eq:c234-gamma}).  
  Based on preliminary STAR data \protect\cite{Luo:2015ewa}.}
\label{fig:k4k2_c234}
\end{figure} 

In Fig. \ref{fig:k3k2_c23} we show $K_{3}/K_{2}$ for different values
of $\gamma _{3}$ in panel (a) and $\gamma _{2}$ in panel (b). 
We observe that, as already discussed before, the preliminary STAR data are
consistent with a constant correlation function in rapidity ($\gamma
_{2}=\gamma _{3}=0$). 
However,  a small positive value of $\gamma_{2}\sim 10^{-2}$ or $ \gamma_{3}\sim 10^{-3}$
would actually improve the agreement slightly. The negative values for
$\gamma_{2}$ and $\gamma_{3}$, on the other hand, appear to be 
disfavored so are large positive values.
The same is true for the comparison with the $K_{4}/K_{2}$ cumulant ratio, which we
show in Fig.~\ref{fig:k4k2_c234}. Again, the data are consistent with  constant rapidity 
correlation functions or perhaps
slightly positive values for $\gamma _{2}$, $\gamma_{3}$, or
$\gamma_{4}$, whereas negative values for $\gamma_{n}$ seem to be
disfavored.\footnote{Specifically we find the following values for $c_{n}^{0}$
  and $\gamma_{n}$ for the blue lines in Figs.~\ref{fig:k3k2_c23} and
  \ref{fig:k4k2_c234}: $\gamma_2 = 10^{-2} $, $c_2^0 \approx
  -2.8\times 10^{-3}$,
$\gamma_3 = 10^{-3} $, $c_3^0 \approx -3.4\times 10^{-4}$, and
$\gamma_4 = 2\times 10^{-4} $, $c_4^0 \approx 3.9\times 10^{-5}$.} 

\begin{figure}[t]
\begin{center}
\includegraphics[scale=0.32]{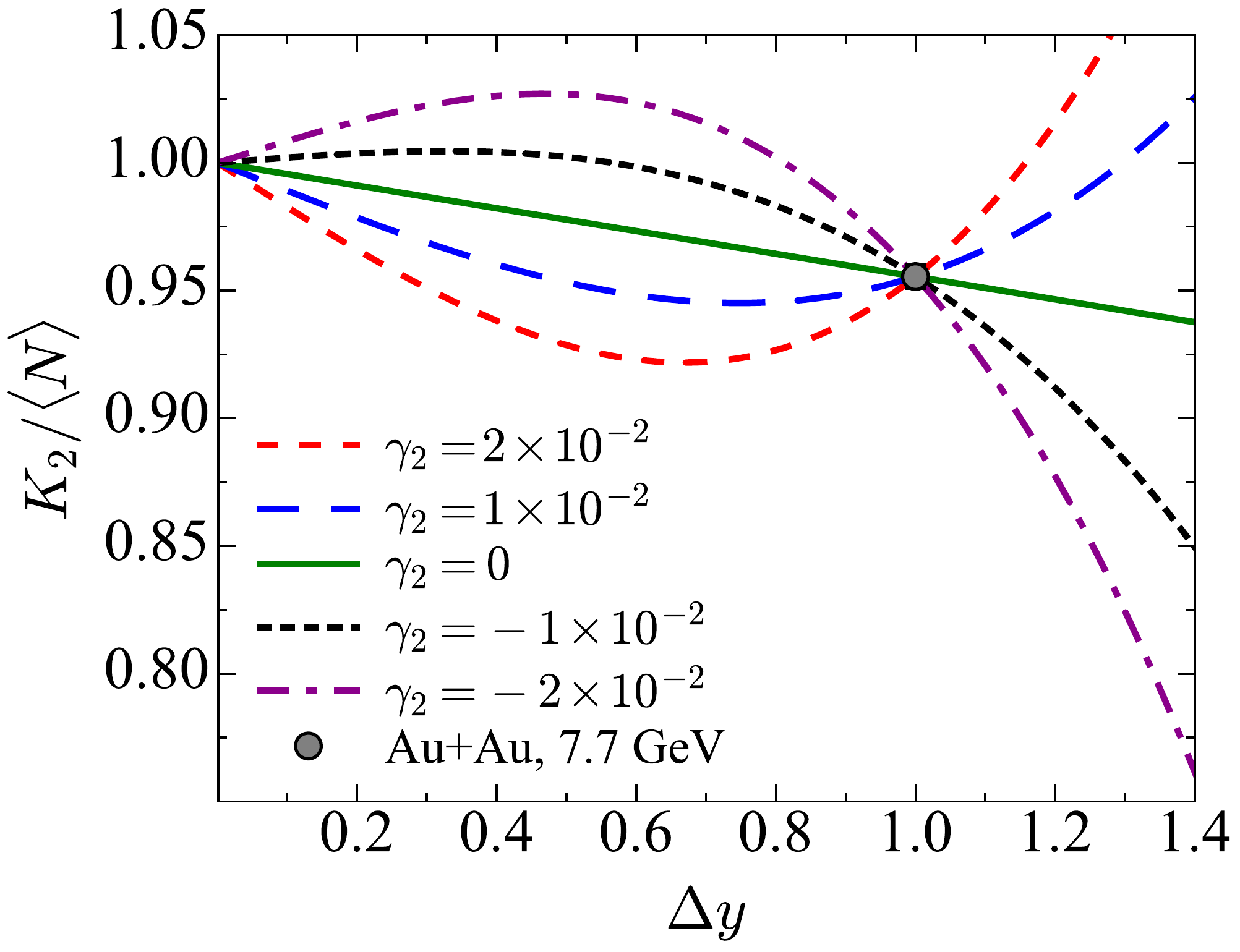} %
\end{center}
\par
\vspace{-5mm}
\caption{The cumulant ratio $K_{2}/K_{1}$ in central Au+Au collisions
  at $\sqrt{s}=7.7$ GeV as a function of the rapidity acceptance $\Delta y$, $|y|<\Delta y/2$, for  
  different values of $\protect\gamma _{2}$.   
  Based on preliminary STAR data \protect\cite{Luo:2015ewa}.}
\label{fig:k2k1_c2}
\end{figure}

Also, the overall picture of slightly ``repulsive''
corrections to the constant correlation functions, i.e., 
$\gamma_{n}\geq 0$ is consistent with the preliminary STAR data
on the two-proton rapidity correlation function, which, as discussed
in the Introduction, 
indicates a peculiar repulsion between protons in rapidity. As these new STAR
measurements only address two proton correlations, the most direct
test would be a comparison of the rapidity dependence of the second
order cumulant or integrated correlation. This is shown in
Fig.~\ref{fig:k2k1_c2}. Unfortunately, at present  there are no data
available for rapidity intervals other than $\Delta y = 1$, and since this
point is used for the determination of the overall constant,
$c_{2}^{0}$, no constraint can be made at this time. However, we wish
to emphasize the strong dependence compared to the size of the error
bar. 
Indeed, the increase in the correlation exhibited in the preliminary
STAR data for the differential correlation functions \cite{star:a1,star:a2} is
consistent with $\gamma_{2}\sim 2\times 10^{-2}$, which would 
correspond to the red dashed curve in
Fig.~\ref{fig:k2k1_c2}. Given the size of
the error bar at $\Delta y = 1$, it should be possible to discriminate
from a constant correlation function, shown by the green solid line.
Needless to say, such a
measurement of the rapidity dependence of
$K_{2}/K_{1}$ would be very valuable to ensure the consistency of the
cumulant measurement with that of the differential correlation
function.\footnote{We note that the preliminary measurements of
  $c_2(y_1,y_2)$ and $K_n$ use different centrality selections, which
  do affect the values of $c_n^0$ and possibly
  $\gamma_n$. Therefore, a direct comparison of the values for
  $\gamma_{2}$ needs to be performed with some care.}

Of course it would be even more valuable to have information about the
differential three- and four-particle correlation functions. 
Therefore, we propose, as a first step, to
measure the rapidity dependence of the couplings, $c_{n}(\Delta
y)$. This will allow for a direct determination of the coefficients,
$\gamma _{n}$, as we demonstrate in Fig.~\ref{fig:cn_dy}, where  we plot 
$c_{n}(\Delta y)/c_{n}^{0}-1$ for $\gamma _{2}=10^{-2}$, $\gamma _{3}=10^{-3}$ and $\gamma _{4}=2\times 10^{-4}$. 
We note that $c_{n}(\Delta y)$ is rather sensitive to $\gamma _{n}$.

In principle it would also be interesting to measure $c_{n}(\Delta y)$ for
higher $n$ such as $n=5$ and $6$. In this case  
\begin{eqnarray}
c_{5}(y_{1},...,y_{5}) &=&c_{5}^{0}+\gamma _{5}\tfrac{1}{10}%
\sum\nolimits_{i,k=1;\text{ }i<k}^{5}\left( y_{i}-y_{k}\right) ^{2},  \notag
\\
c_{6}(y_{1},...,y_{6}) &=&c_{6}^{0}+\gamma _{6}\tfrac{1}{15}%
\sum\nolimits_{i,k=1;\text{ }i<k}^{6}\left( y_{i}-y_{k}\right) ^{2},
\end{eqnarray}%
and $c_{n}(\Delta y)$ is given by Eq. (\ref{eq:cn-delta-y}).

\section{Discussion and conclusions}
\label{Sec:Conclusion}

\begin{figure}[t!]
\begin{center}
\includegraphics[scale=0.32]{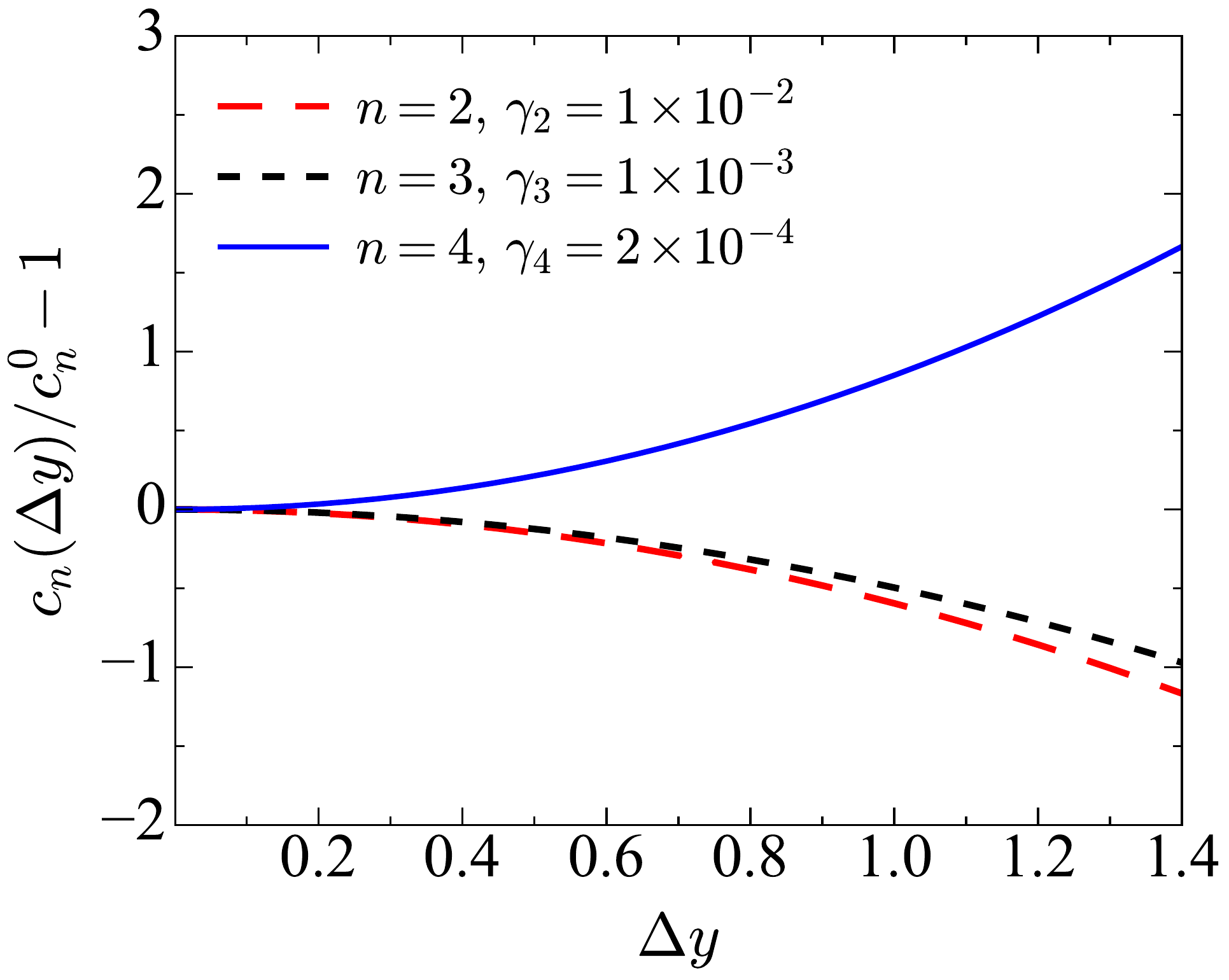}
\end{center}
\par
\vspace{-5mm}
\caption{The ratio of the couplings $c_{n}(\Delta y)/c_{n}^{0}-1$, see
         Eqs.~(\ref{eq:coupling},\ref{eq:cn-delta-y}) for $\gamma _{2}=10^{-2}$, 
         $\gamma _{3}=10^{-3}$ and $\gamma _{4}=2\times 10^{-4}$. 
         $\Delta y$ denotes the size of the rapidity window, $|y|<\Delta y/2$, 
         which the reduced correlation functions are integrated over.}
\label{fig:cn_dy}
\end{figure}

Before we conclude let us discuss the main findings of this paper.
\begin{itemize}
\item The preliminary data for the proton cumulant ratio $K_{4}/K_{2}$ obtained by the
  STAR collaboration at $\sqrt{s}=7.7$ GeV are consistent with long-range correlations in
  both rapidity and transverse momentum. As a result the cumulants
  effectively depend only on the number of protons $\ave{N}$ in the
  acceptance. Therefore, we predict that new measurements with increased
  acceptance will lead to even larger values for the
  $K_{4}/K_{2}$. Naturally this increase will be limited eventually by
  global charge conservation as discussed in
  \cite{Bzdak:2012an}, and the ansatz for the correlation function,
  Eq.~\eqref{eq:cn-const} will have its limitation for large $\Delta
  y$. Consequently our present prediction for large $\Delta y > 1$
  needs  to be taken with a grain of salt. 
\item Allowing for small deviation from a constant value we find that
  a slightly ``repulsive'' rapidity dependence is favored by
  the data. By ``repulsive'' we mean that the correlation function
  increases with increasing rapidity separation between protons. Or in
  other words, we find that $\gamma_{n }>0 $ in
  Eq.~\eqref{eq:c234-gamma} is favored.  Perhaps this may be the first
  evidence for ``repulsive'' three-proton and four-proton correlations.
\item Clearly, as demonstrated in Fig.~\ref{fig:cn_dy}, a measurement
  of the couplings as a function of the rapidity and transverse
  momentum windows would be very valuable to shed more light on the
  ranges and detailed shapes of the correlation functions.
\item Finally  we want to reiterate that the fact that cumulant ratios
  for small acceptance or, more precisely for a small number of
  particles, are close to unity does not necessarily imply the absence of correlations. 
  This is demonstrated in Fig.~\ref{fig:k4k2_scaling} where we actually assume
  a \textit{constant} correlation. In addition, this may also be the
  reason that anti-protons show a cumulant ratio of $K_{4}/K_{2}\simeq
  1$ at low energies whereas the protons show a significant deviation
  from unity. \\
  We further demonstrated that  global baryon conservation
    fully determines the cumulant ratios, integrated correlation
    functions, and couplings close to the full acceptance
    regardless of any additional dynamics. In addition we showed that, as a result of
    baryon number conservation, the
    cumulants in a given phase-space window and their complements are closely related, 
    see Eq. (\ref{eq:relation}) and the Appendix.
\end{itemize}

To summarize, we have studied the rapidity dependence of cumulants, integrated
correlation functions and couplings based on the presently available
preliminary STAR collaboration data \cite{Luo:2015ewa}. Although we found that, within
the present experimental errors the data are consistent with rapidity
independent correlations, a slightly ``repulsive''  component seems to
be favored. This would be consistent with the preliminary measurement
of two-particle differential proton correlations by STAR
\cite{star:a1,star:a2}. To gain further insight, in particular, into the three-proton
and four-proton correlations, we proposed to measure the dependence
of the couplings as a function of the rapidity window.

\bigskip

 \begin{acknowledgements}
We thank the STAR collaboration for providing us with their
preliminary data. We further thank A.~Bialas and W.~Llope for stimulating discussions.
A.B. was supported by the Ministry of Science and Higher Education (MNiSW) and 
by the National Science Centre, Grant No. DEC-2014/15/B/ST2/00175, and 
in part by DEC-2013/09/B/ST2/00497.
V.K. was supported by the Office of Nuclear Physics in the US Department of Energy's Office 
of Science under Contract No. DE-AC02-05CH11231.
 \end{acknowledgements}
 
\appendix

\section{Full acceptance}

Suppose we divide the full phase space into the two, not
necessarily equal sized regions denoted by the subscripts $(a)$ and $(b)$. 
Let $P_{(a)}(N_{(a)})$ be the probability to observe $N_{(a)}$ baryons in the
phase space region $(a)$. 
The probability to have $N_{(b)}$ baryons in the remaining part
of the entire phase space, $P_{(b)}(N_{(b)})$, is given by 
$P_{(b)}(N_{(b)})=P_{(a)}(N_{(a)})=P_{(a)}(B-N_{(b)})$ since
$N_{(a)} = B - N_{(b)}$, where $B$ is the conserved number of
baryons. Here we assume that we can ignore anti-baryons. 
The cumulant generating function for the phase space region $(a)$,
$h_{(a)}(t)$ is given by
\begin{eqnarray}
h_{(a)}(t) &=&\log \left[ \sum\nolimits_{N_{(a)}}P_{(a)}(N_{(a)})e^{N_{(a)}t}\right] \notag \\
&=&\log \left[ \sum\nolimits_{N_{(b)}}P_{(a)}(B-N_{(b)})e^{(B-N_{(b)})t}\right] \notag \\
&=&\log \left[ \sum\nolimits_{N_{(b)}}P_{(b)}(N_{(b)})e^{(B-N_{(b)})t}\right]  \notag \\
&=&h_{(b)}(-t)+Bt ,
\end{eqnarray}
where $h_{(b)}(t)$ is the cumulant generating function for phase space
region $(b)$.
The cumulants in the two regions, $(a)$ and $(b)$, are given by the derivatives at $t=0$,
\begin{equation}
K_{n,(a)}=\frac{d^{n}}{dt^{n}}h_{(a)}(t)|_{t=0}, \quad K_{n,(b)}=\frac{d^{n}}{dt^{n}}h_{(b)}(t)|_{t=0}.
\end{equation}%
Thus we get for $n=1$
\begin{equation}
\ave{N_{(a)}}=K_{1,(a)}=B-K_{1,(b)}=B-\ave{N_{(b)}},
\end{equation}
and for $n \geq 2$ 
\begin{equation}
K_{n,(a)}=(-1)^{n}K_{n,(b)}.
\label{eq:cum_symmetry}
\end{equation}. 

Given this relation between the cumulants of the two regions and using
Eqs.~\eqref{eq:K1234}-\eqref{eq:K4C} we also can find the relation
between the integrated correlation functions $C_{{n,(a)}}$ and $C_{n,(b)}$
in regions $(a)$ and $(b)$, respectively.  
\begin{eqnarray}
C_{2,(a)} &=& -B + 2 \langle N_{(b)} \rangle + C_{2,(b)} , \non
C_{3,(a)} &=& 2 B - 6 \langle N_{(b)} \rangle - 6 C_{2,(b)} - C_{3,(b)} , \non
C_{4,(a)} &=& -6 B +  24 \langle N_{(b)} \rangle + 36 C_{2,(b)}+12 C_{3,(b)}+C_{4,(b)} .
\end{eqnarray}

Clearly, the integrated correlation functions do not show any symmetry
between the two complementary regions of the
phase space. The same is also true for the couplings $c_{n}$. In the
limit where $\ave{N_{(a)}} \rightarrow B$ and thus $\ave{N_{(b)}} \rightarrow 0$ 
we find, following the above equations, that
$C_{2,(a)} \rightarrow -B$,  $C_{3,(a)} \rightarrow 2 B$, and $C_{4,(a)}
\rightarrow -6 B$. In this case, the couplings become $c_{2,(a)}\rightarrow
-\frac{1}{B}$, $c_{3,(a)} \rightarrow \frac{2}{B^{2}}$, and
$c_{4,(a)}\rightarrow -\frac{6}{B^{3}}$, and again are
determined entirely by the total baryon number $B$.
For the complementary region $(b)$,
on the other hand, we have the limit of $\ave{N_{b}} \rightarrow 0$,
in which case, as discussed in Section~\ref{sec:poisson}, dynamics
beyond baryon number conservation
also affects the couplings.


\begin{thebibliography}{53}%
\makeatletter
\providecommand \@ifxundefined [1]{%
 \@ifx{#1\undefined}
}%
\providecommand \@ifnum [1]{%
 \ifnum #1\expandafter \@firstoftwo
 \else \expandafter \@secondoftwo
 \fi
}%
\providecommand \@ifx [1]{%
 \ifx #1\expandafter \@firstoftwo
 \else \expandafter \@secondoftwo
 \fi
}%
\providecommand \natexlab [1]{#1}%
\providecommand \enquote  [1]{``#1''}%
\providecommand \bibnamefont  [1]{#1}%
\providecommand \bibfnamefont [1]{#1}%
\providecommand \citenamefont [1]{#1}%
\providecommand \href@noop [0]{\@secondoftwo}%
\providecommand \href [0]{\begingroup \@sanitize@url \@href}%
\providecommand \@href[1]{\@@startlink{#1}\@@href}%
\providecommand \@@href[1]{\endgroup#1\@@endlink}%
\providecommand \@sanitize@url [0]{\catcode `\\12\catcode `\$12\catcode
  `\&12\catcode `\#12\catcode `\^12\catcode `\_12\catcode `\%12\relax}%
\providecommand \@@startlink[1]{}%
\providecommand \@@endlink[0]{}%
\providecommand \url  [0]{\begingroup\@sanitize@url \@url }%
\providecommand \@url [1]{\endgroup\@href {#1}{\urlprefix }}%
\providecommand \urlprefix  [0]{URL }%
\providecommand \Eprint [0]{\href }%
\providecommand \doibase [0]{http://dx.doi.org/}%
\providecommand \selectlanguage [0]{\@gobble}%
\providecommand \bibinfo  [0]{\@secondoftwo}%
\providecommand \bibfield  [0]{\@secondoftwo}%
\providecommand \translation [1]{[#1]}%
\providecommand \BibitemOpen [0]{}%
\providecommand \bibitemStop [0]{}%
\providecommand \bibitemNoStop [0]{.\EOS\space}%
\providecommand \EOS [0]{\spacefactor3000\relax}%
\providecommand \BibitemShut  [1]{\csname bibitem#1\endcsname}%
\let\auto@bib@innerbib\@empty
\bibitem [{\citenamefont {Stephanov}\ \emph {et~al.}(1998)\citenamefont
  {Stephanov}, \citenamefont {Rajagopal},\ and\ \citenamefont
  {Shuryak}}]{Stephanov:1998dy}%
  \BibitemOpen
  \bibfield  {author} {\bibinfo {author} {\bibfnamefont {M.~A.}\ \bibnamefont
  {Stephanov}}, \bibinfo {author} {\bibfnamefont {K.}~\bibnamefont
  {Rajagopal}}, \ and\ \bibinfo {author} {\bibfnamefont {E.~V.}\ \bibnamefont
  {Shuryak}},\ }\href@noop {} {\bibfield  {journal} {\bibinfo  {journal} {Phys.
  Rev. Lett.}\ }\textbf {\bibinfo {volume} {81}},\ \bibinfo {pages} {4816}
  (\bibinfo {year} {1998})},\ \Eprint {http://arXiv.org/abs/hep-ph/9806219}
  {hep-ph/9806219} \BibitemShut {NoStop}%
\bibitem [{\citenamefont {Stephanov}(2009)}]{Stephanov:2008qz}%
  \BibitemOpen
  \bibfield  {author} {\bibinfo {author} {\bibfnamefont {M.}~\bibnamefont
  {Stephanov}},\ }\href {\doibase 10.1103/PhysRevLett.102.032301} {\bibfield
  {journal} {\bibinfo  {journal} {Phys.Rev.Lett.}\ }\textbf {\bibinfo {volume}
  {102}},\ \bibinfo {pages} {032301} (\bibinfo {year} {2009})},\ \Eprint
  {http://arxiv.org/abs/0809.3450} {arXiv:0809.3450 [hep-ph]} \BibitemShut
  {NoStop}%
\bibitem [{\citenamefont {Skokov}\ \emph {et~al.}(2011)\citenamefont {Skokov},
  \citenamefont {Friman},\ and\ \citenamefont {Redlich}}]{Skokov:2010uh}%
  \BibitemOpen
  \bibfield  {author} {\bibinfo {author} {\bibfnamefont {V.}~\bibnamefont
  {Skokov}}, \bibinfo {author} {\bibfnamefont {B.}~\bibnamefont {Friman}}, \
  and\ \bibinfo {author} {\bibfnamefont {K.}~\bibnamefont {Redlich}},\ }\href
  {\doibase 10.1103/PhysRevC.83.054904} {\bibfield  {journal} {\bibinfo
  {journal} {Phys.Rev.}\ }\textbf {\bibinfo {volume} {C83}},\ \bibinfo {pages}
  {054904} (\bibinfo {year} {2011})},\ \Eprint {http://arxiv.org/abs/1008.4570}
  {arXiv:1008.4570 [hep-ph]} \BibitemShut {NoStop}%
\bibitem [{\citenamefont {Friman}\ \emph {et~al.}(2011)\citenamefont {Friman},
  \citenamefont {Karsch}, \citenamefont {Redlich},\ and\ \citenamefont
  {Skokov}}]{Friman:2011pf}%
  \BibitemOpen
  \bibfield  {author} {\bibinfo {author} {\bibfnamefont {B.}~\bibnamefont
  {Friman}}, \bibinfo {author} {\bibfnamefont {F.}~\bibnamefont {Karsch}},
  \bibinfo {author} {\bibfnamefont {K.}~\bibnamefont {Redlich}}, \ and\
  \bibinfo {author} {\bibfnamefont {V.}~\bibnamefont {Skokov}},\ }\href
  {\doibase 10.1140/epjc/s10052-011-1694-2} {\bibfield  {journal} {\bibinfo
  {journal} {Eur.Phys.J.}\ }\textbf {\bibinfo {volume} {C71}},\ \bibinfo
  {pages} {1694} (\bibinfo {year} {2011})},\ \Eprint
  {http://arxiv.org/abs/1103.3511} {arXiv:1103.3511 [hep-ph]} \BibitemShut
  {NoStop}%
\bibitem [{\citenamefont {Adamczyk}\ \emph
  {et~al.}(2014{\natexlab{a}})\citenamefont {Adamczyk} \emph
  {et~al.}}]{Adamczyk:2013dal}%
  \BibitemOpen
  \bibfield  {author} {\bibinfo {author} {\bibfnamefont {L.}~\bibnamefont
  {Adamczyk}} \emph {et~al.} (\bibinfo {collaboration} {STAR}),\ }\href
  {\doibase 10.1103/PhysRevLett.112.032302} {\bibfield  {journal} {\bibinfo
  {journal} {Phys. Rev. Lett.}\ }\textbf {\bibinfo {volume} {112}},\ \bibinfo
  {pages} {032302} (\bibinfo {year} {2014}{\natexlab{a}})},\ \Eprint
  {http://arxiv.org/abs/1309.5681} {arXiv:1309.5681 [nucl-ex]} \BibitemShut
  {NoStop}%
\bibitem [{\citenamefont {Adamczyk}\ \emph
  {et~al.}(2014{\natexlab{b}})\citenamefont {Adamczyk} \emph
  {et~al.}}]{Adamczyk:2014fia}%
  \BibitemOpen
  \bibfield  {author} {\bibinfo {author} {\bibfnamefont {L.}~\bibnamefont
  {Adamczyk}} \emph {et~al.} (\bibinfo {collaboration} {STAR}),\ }\href
  {\doibase 10.1103/PhysRevLett.113.092301} {\bibfield  {journal} {\bibinfo
  {journal} {Phys. Rev. Lett.}\ }\textbf {\bibinfo {volume} {113}},\ \bibinfo
  {pages} {092301} (\bibinfo {year} {2014}{\natexlab{b}})},\ \Eprint
  {http://arxiv.org/abs/1402.1558} {arXiv:1402.1558 [nucl-ex]} \BibitemShut
  {NoStop}%
\bibitem [{\citenamefont {Borsanyi}\ \emph {et~al.}(2012)\citenamefont
  {Borsanyi}, \citenamefont {Fodor}, \citenamefont {Katz}, \citenamefont
  {Krieg}, \citenamefont {Ratti} \emph {et~al.}}]{Borsanyi:2011sw}%
  \BibitemOpen
  \bibfield  {author} {\bibinfo {author} {\bibfnamefont {S.}~\bibnamefont
  {Borsanyi}}, \bibinfo {author} {\bibfnamefont {Z.}~\bibnamefont {Fodor}},
  \bibinfo {author} {\bibfnamefont {S.~D.}\ \bibnamefont {Katz}}, \bibinfo
  {author} {\bibfnamefont {S.}~\bibnamefont {Krieg}}, \bibinfo {author}
  {\bibfnamefont {C.}~\bibnamefont {Ratti}},  \emph {et~al.},\ }\href {\doibase
  10.1007/JHEP01(2012)138} {\bibfield  {journal} {\bibinfo  {journal} {JHEP}\
  }\textbf {\bibinfo {volume} {1201}},\ \bibinfo {pages} {138} (\bibinfo {year}
  {2012})},\ \Eprint {http://arxiv.org/abs/1112.4416} {arXiv:1112.4416
  [hep-lat]} \BibitemShut {NoStop}%
\bibitem [{\citenamefont {Bazavov}\ \emph {et~al.}(2012)\citenamefont {Bazavov}
  \emph {et~al.}}]{Bazavov:2012jq}%
  \BibitemOpen
  \bibfield  {author} {\bibinfo {author} {\bibfnamefont {A.}~\bibnamefont
  {Bazavov}} \emph {et~al.} (\bibinfo {collaboration} {HotQCD Collaboration}),\
  }\href {\doibase 10.1103/PhysRevD.86.034509} {\bibfield  {journal} {\bibinfo
  {journal} {Phys.Rev.}\ }\textbf {\bibinfo {volume} {D86}},\ \bibinfo {pages}
  {034509} (\bibinfo {year} {2012})},\ \Eprint {http://arxiv.org/abs/1203.0784}
  {arXiv:1203.0784 [hep-lat]} \BibitemShut {NoStop}%
\bibitem [{\citenamefont {Bellwied}\ \emph {et~al.}(2015)\citenamefont
  {Bellwied}, \citenamefont {Borsanyi}, \citenamefont {Fodor}, \citenamefont
  {Katz}, \citenamefont {Pasztor}, \citenamefont {Ratti},\ and\ \citenamefont
  {Szabo}}]{Bellwied:2015lba}%
  \BibitemOpen
  \bibfield  {author} {\bibinfo {author} {\bibfnamefont {R.}~\bibnamefont
  {Bellwied}}, \bibinfo {author} {\bibfnamefont {S.}~\bibnamefont {Borsanyi}},
  \bibinfo {author} {\bibfnamefont {Z.}~\bibnamefont {Fodor}}, \bibinfo
  {author} {\bibfnamefont {S.~D.}\ \bibnamefont {Katz}}, \bibinfo {author}
  {\bibfnamefont {A.}~\bibnamefont {Pasztor}}, \bibinfo {author} {\bibfnamefont
  {C.}~\bibnamefont {Ratti}}, \ and\ \bibinfo {author} {\bibfnamefont {K.~K.}\
  \bibnamefont {Szabo}},\ }\href {\doibase 10.1103/PhysRevD.92.114505}
  {\bibfield  {journal} {\bibinfo  {journal} {Phys. Rev.}\ }\textbf {\bibinfo
  {volume} {D92}},\ \bibinfo {pages} {114505} (\bibinfo {year} {2015})},\
  \Eprint {http://arxiv.org/abs/1507.04627} {arXiv:1507.04627 [hep-lat]}
  \BibitemShut {NoStop}%
\bibitem [{\citenamefont {Gavai}\ and\ \citenamefont
  {Gupta}(2011)}]{Gavai:2010zn}%
  \BibitemOpen
  \bibfield  {author} {\bibinfo {author} {\bibfnamefont {R.~V.}\ \bibnamefont
  {Gavai}}\ and\ \bibinfo {author} {\bibfnamefont {S.}~\bibnamefont {Gupta}},\
  }\href {\doibase 10.1016/j.physletb.2011.01.006} {\bibfield  {journal}
  {\bibinfo  {journal} {Phys. Lett.}\ }\textbf {\bibinfo {volume} {B696}},\
  \bibinfo {pages} {459} (\bibinfo {year} {2011})},\ \Eprint
  {http://arxiv.org/abs/1001.3796} {arXiv:1001.3796 [hep-lat]} \BibitemShut
  {NoStop}%
\bibitem [{\citenamefont {Kitazawa}\ \emph {et~al.}(2014)\citenamefont
  {Kitazawa}, \citenamefont {Asakawa},\ and\ \citenamefont
  {Ono}}]{Kitazawa:2013bta}%
  \BibitemOpen
  \bibfield  {author} {\bibinfo {author} {\bibfnamefont {M.}~\bibnamefont
  {Kitazawa}}, \bibinfo {author} {\bibfnamefont {M.}~\bibnamefont {Asakawa}}, \
  and\ \bibinfo {author} {\bibfnamefont {H.}~\bibnamefont {Ono}},\ }\href
  {\doibase 10.1016/j.physletb.2013.12.008} {\bibfield  {journal} {\bibinfo
  {journal} {Phys.Lett.}\ }\textbf {\bibinfo {volume} {B728}},\ \bibinfo
  {pages} {386} (\bibinfo {year} {2014})},\ \Eprint
  {http://arxiv.org/abs/1307.2978} {arXiv:1307.2978 [nucl-th]} \BibitemShut
  {NoStop}%
\bibitem [{\citenamefont {Asakawa}\ and\ \citenamefont
  {Kitazawa}(2016)}]{Asakawa:2015ybt}%
  \BibitemOpen
  \bibfield  {author} {\bibinfo {author} {\bibfnamefont {M.}~\bibnamefont
  {Asakawa}}\ and\ \bibinfo {author} {\bibfnamefont {M.}~\bibnamefont
  {Kitazawa}},\ }\href {\doibase 10.1016/j.ppnp.2016.04.002} {\bibfield
  {journal} {\bibinfo  {journal} {Prog. Part. Nucl. Phys.}\ }\textbf {\bibinfo
  {volume} {90}},\ \bibinfo {pages} {299} (\bibinfo {year} {2016})},\ \Eprint
  {http://arxiv.org/abs/1512.05038} {arXiv:1512.05038 [nucl-th]} \BibitemShut
  {NoStop}%
\bibitem [{\citenamefont {Mukherjee}\ \emph {et~al.}(2016)\citenamefont
  {Mukherjee}, \citenamefont {Steinheimer},\ and\ \citenamefont
  {Schramm}}]{Mukherjee:2016nhb}%
  \BibitemOpen
  \bibfield  {author} {\bibinfo {author} {\bibfnamefont {A.}~\bibnamefont
  {Mukherjee}}, \bibinfo {author} {\bibfnamefont {J.}~\bibnamefont
  {Steinheimer}}, \ and\ \bibinfo {author} {\bibfnamefont {S.}~\bibnamefont
  {Schramm}},\ }\href@noop {} {\  (\bibinfo {year} {2016})},\ \Eprint
  {http://arxiv.org/abs/1611.10144} {arXiv:1611.10144 [nucl-th]} \BibitemShut
  {NoStop}%
\bibitem [{\citenamefont {Herold}\ \emph {et~al.}(2016)\citenamefont {Herold},
  \citenamefont {Nahrgang}, \citenamefont {Yan},\ and\ \citenamefont
  {Kobdaj}}]{Herold:2016uvv}%
  \BibitemOpen
  \bibfield  {author} {\bibinfo {author} {\bibfnamefont {C.}~\bibnamefont
  {Herold}}, \bibinfo {author} {\bibfnamefont {M.}~\bibnamefont {Nahrgang}},
  \bibinfo {author} {\bibfnamefont {Y.}~\bibnamefont {Yan}}, \ and\ \bibinfo
  {author} {\bibfnamefont {C.}~\bibnamefont {Kobdaj}},\ }\href {\doibase
  10.1103/PhysRevC.93.021902} {\bibfield  {journal} {\bibinfo  {journal} {Phys.
  Rev.}\ }\textbf {\bibinfo {volume} {C93}},\ \bibinfo {pages} {021902}
  (\bibinfo {year} {2016})},\ \Eprint {http://arxiv.org/abs/1601.04839}
  {arXiv:1601.04839 [hep-ph]} \BibitemShut {NoStop}%
\bibitem [{\citenamefont {Chatterjee}\ \emph {et~al.}(2016)\citenamefont
  {Chatterjee}, \citenamefont {Chatterjee}, \citenamefont {Nayak},\ and\
  \citenamefont {Sahoo}}]{Chatterjee:2016mve}%
  \BibitemOpen
  \bibfield  {author} {\bibinfo {author} {\bibfnamefont {A.}~\bibnamefont
  {Chatterjee}}, \bibinfo {author} {\bibfnamefont {S.}~\bibnamefont
  {Chatterjee}}, \bibinfo {author} {\bibfnamefont {T.~K.}\ \bibnamefont
  {Nayak}}, \ and\ \bibinfo {author} {\bibfnamefont {N.~R.}\ \bibnamefont
  {Sahoo}},\ }\href {\doibase 10.1088/0954-3899/43/12/125103} {\bibfield
  {journal} {\bibinfo  {journal} {J. Phys.}\ }\textbf {\bibinfo {volume}
  {G43}},\ \bibinfo {pages} {125103} (\bibinfo {year} {2016})},\ \Eprint
  {http://arxiv.org/abs/1606.09573} {arXiv:1606.09573 [nucl-ex]} \BibitemShut
  {NoStop}%
\bibitem [{\citenamefont {Lacey}\ \emph {et~al.}(2016)\citenamefont {Lacey},
  \citenamefont {Liu}, \citenamefont {Magdy}, \citenamefont {Schweid},\ and\
  \citenamefont {Ajitanand}}]{Lacey:2016tsw}%
  \BibitemOpen
  \bibfield  {author} {\bibinfo {author} {\bibfnamefont {R.~A.}\ \bibnamefont
  {Lacey}}, \bibinfo {author} {\bibfnamefont {P.}~\bibnamefont {Liu}}, \bibinfo
  {author} {\bibfnamefont {N.}~\bibnamefont {Magdy}}, \bibinfo {author}
  {\bibfnamefont {B.}~\bibnamefont {Schweid}}, \ and\ \bibinfo {author}
  {\bibfnamefont {N.~N.}\ \bibnamefont {Ajitanand}},\ }\href@noop {} {\
  (\bibinfo {year} {2016})},\ \Eprint {http://arxiv.org/abs/1606.08071}
  {arXiv:1606.08071 [nucl-ex]} \BibitemShut {NoStop}%
\bibitem [{\citenamefont {Hippert}\ and\ \citenamefont
  {Fraga}(2017)}]{Hippert:2017xoj}%
  \BibitemOpen
  \bibfield  {author} {\bibinfo {author} {\bibfnamefont {M.}~\bibnamefont
  {Hippert}}\ and\ \bibinfo {author} {\bibfnamefont {E.~S.}\ \bibnamefont
  {Fraga}},\ }\href@noop {} {\  (\bibinfo {year} {2017})},\ \Eprint
  {http://arxiv.org/abs/1702.02028} {arXiv:1702.02028 [hep-ph]} \BibitemShut
  {NoStop}%
\bibitem [{\citenamefont {Rougemont}\ \emph {et~al.}(2017)\citenamefont
  {Rougemont}, \citenamefont {Critelli}, \citenamefont {Noronha-Hostler},
  \citenamefont {Noronha},\ and\ \citenamefont {Ratti}}]{Rougemont:2017tlu}%
  \BibitemOpen
  \bibfield  {author} {\bibinfo {author} {\bibfnamefont {R.}~\bibnamefont
  {Rougemont}}, \bibinfo {author} {\bibfnamefont {R.}~\bibnamefont {Critelli}},
  \bibinfo {author} {\bibfnamefont {J.}~\bibnamefont {Noronha-Hostler}},
  \bibinfo {author} {\bibfnamefont {J.}~\bibnamefont {Noronha}}, \ and\
  \bibinfo {author} {\bibfnamefont {C.}~\bibnamefont {Ratti}},\ }\href@noop {}
  {\  (\bibinfo {year} {2017})},\ \Eprint {http://arxiv.org/abs/1704.05558}
  {arXiv:1704.05558 [hep-ph]} \BibitemShut {NoStop}%
\bibitem [{\citenamefont {Almasi}\ \emph {et~al.}(2017)\citenamefont {Almasi},
  \citenamefont {Friman},\ and\ \citenamefont {Redlich}}]{Almasi:2017bhq}%
  \BibitemOpen
  \bibfield  {author} {\bibinfo {author} {\bibfnamefont {G.~A.}\ \bibnamefont
  {Almasi}}, \bibinfo {author} {\bibfnamefont {B.}~\bibnamefont {Friman}}, \
  and\ \bibinfo {author} {\bibfnamefont {K.}~\bibnamefont {Redlich}},\
  }\href@noop {} {\  (\bibinfo {year} {2017})},\ \Eprint
  {http://arxiv.org/abs/1703.05947} {arXiv:1703.05947 [hep-ph]} \BibitemShut
  {NoStop}%
\bibitem [{\citenamefont {He}\ and\ \citenamefont {Luo}(2017)}]{He:2017zpg}%
  \BibitemOpen
  \bibfield  {author} {\bibinfo {author} {\bibfnamefont {S.}~\bibnamefont
  {He}}\ and\ \bibinfo {author} {\bibfnamefont {X.}~\bibnamefont {Luo}},\
  }\href@noop {} {\  (\bibinfo {year} {2017})},\ \Eprint
  {http://arxiv.org/abs/1704.00423} {arXiv:1704.00423 [nucl-ex]} \BibitemShut
  {NoStop}%
\bibitem [{\citenamefont {Koch}(2010)}]{Koch:2008ia}%
  \BibitemOpen
  \bibfield  {author} {\bibinfo {author} {\bibfnamefont {V.}~\bibnamefont
  {Koch}},\ }in\ \href {\doibase 10.1007/978-3-642-01539-7_20} {\emph {\bibinfo
  {booktitle} {Relativistic Heavy Ion Physics}}},\ \bibinfo {series}
  {Landolt-Boernstein New Series I}, Vol.~\bibinfo {volume} {23},\ \bibinfo
  {editor} {edited by\ \bibinfo {editor} {\bibfnamefont {R.}~\bibnamefont
  {Stock}}}\ (\bibinfo  {publisher} {Springer},\ \bibinfo {address}
  {Heidelberg},\ \bibinfo {year} {2010})\ pp.\ \bibinfo {pages} {626--652},\
  \Eprint {http://arxiv.org/abs/0810.2520} {arXiv:0810.2520 [nucl-th]}
  \BibitemShut {NoStop}%
\bibitem [{\citenamefont {Bzdak}\ \emph {et~al.}(2013)\citenamefont {Bzdak},
  \citenamefont {Koch},\ and\ \citenamefont {Skokov}}]{Bzdak:2012an}%
  \BibitemOpen
  \bibfield  {author} {\bibinfo {author} {\bibfnamefont {A.}~\bibnamefont
  {Bzdak}}, \bibinfo {author} {\bibfnamefont {V.}~\bibnamefont {Koch}}, \ and\
  \bibinfo {author} {\bibfnamefont {V.}~\bibnamefont {Skokov}},\ }\href
  {\doibase 10.1103/PhysRevC.87.014901} {\bibfield  {journal} {\bibinfo
  {journal} {Phys. Rev.}\ }\textbf {\bibinfo {volume} {C87}},\ \bibinfo {pages}
  {014901} (\bibinfo {year} {2013})},\ \Eprint {http://arxiv.org/abs/1203.4529}
  {arXiv:1203.4529 [hep-ph]} \BibitemShut {NoStop}%
\bibitem [{\citenamefont {Skokov}\ \emph {et~al.}(2013)\citenamefont {Skokov},
  \citenamefont {Friman},\ and\ \citenamefont {Redlich}}]{Skokov:2012ds}%
  \BibitemOpen
  \bibfield  {author} {\bibinfo {author} {\bibfnamefont {V.}~\bibnamefont
  {Skokov}}, \bibinfo {author} {\bibfnamefont {B.}~\bibnamefont {Friman}}, \
  and\ \bibinfo {author} {\bibfnamefont {K.}~\bibnamefont {Redlich}},\ }\href
  {\doibase 10.1103/PhysRevC.88.034911} {\bibfield  {journal} {\bibinfo
  {journal} {Phys. Rev.}\ }\textbf {\bibinfo {volume} {C88}},\ \bibinfo {pages}
  {034911} (\bibinfo {year} {2013})},\ \Eprint {http://arxiv.org/abs/1205.4756}
  {arXiv:1205.4756 [hep-ph]} \BibitemShut {NoStop}%
\bibitem [{\citenamefont {Bzdak}\ and\ \citenamefont
  {Koch}(2012)}]{Bzdak:2012ab}%
  \BibitemOpen
  \bibfield  {author} {\bibinfo {author} {\bibfnamefont {A.}~\bibnamefont
  {Bzdak}}\ and\ \bibinfo {author} {\bibfnamefont {V.}~\bibnamefont {Koch}},\
  }\href {\doibase 10.1103/PhysRevC.86.044904} {\bibfield  {journal} {\bibinfo
  {journal} {Phys. Rev.}\ }\textbf {\bibinfo {volume} {C86}},\ \bibinfo {pages}
  {044904} (\bibinfo {year} {2012})},\ \Eprint {http://arxiv.org/abs/1206.4286}
  {arXiv:1206.4286 [nucl-th]} \BibitemShut {NoStop}%
\bibitem [{\citenamefont {Bzdak}\ and\ \citenamefont
  {Koch}(2015)}]{Bzdak:2013pha}%
  \BibitemOpen
  \bibfield  {author} {\bibinfo {author} {\bibfnamefont {A.}~\bibnamefont
  {Bzdak}}\ and\ \bibinfo {author} {\bibfnamefont {V.}~\bibnamefont {Koch}},\
  }\href {\doibase 10.1103/PhysRevC.91.027901} {\bibfield  {journal} {\bibinfo
  {journal} {Phys. Rev.}\ }\textbf {\bibinfo {volume} {C91}},\ \bibinfo {pages}
  {027901} (\bibinfo {year} {2015})},\ \Eprint {http://arxiv.org/abs/1312.4574}
  {arXiv:1312.4574 [nucl-th]} \BibitemShut {NoStop}%
\bibitem [{\citenamefont {Luo}(2015{\natexlab{a}})}]{Luo:2014rea}%
  \BibitemOpen
  \bibfield  {author} {\bibinfo {author} {\bibfnamefont {X.}~\bibnamefont
  {Luo}},\ }\href {\doibase 10.1103/PhysRevC.91.034907} {\bibfield  {journal}
  {\bibinfo  {journal} {Phys. Rev.}\ }\textbf {\bibinfo {volume} {C91}},\
  \bibinfo {pages} {034907} (\bibinfo {year} {2015}{\natexlab{a}})},\ \Eprint
  {http://arxiv.org/abs/1410.3914} {arXiv:1410.3914 [physics.data-an]}
  \BibitemShut {NoStop}%
\bibitem [{\citenamefont {Nonaka}\ \emph {et~al.}(2016)\citenamefont {Nonaka},
  \citenamefont {Sugiura}, \citenamefont {Esumi}, \citenamefont {Masui},\ and\
  \citenamefont {Luo}}]{Nonaka:2016xje}%
  \BibitemOpen
  \bibfield  {author} {\bibinfo {author} {\bibfnamefont {T.}~\bibnamefont
  {Nonaka}}, \bibinfo {author} {\bibfnamefont {T.}~\bibnamefont {Sugiura}},
  \bibinfo {author} {\bibfnamefont {S.}~\bibnamefont {Esumi}}, \bibinfo
  {author} {\bibfnamefont {H.}~\bibnamefont {Masui}}, \ and\ \bibinfo {author}
  {\bibfnamefont {X.}~\bibnamefont {Luo}},\ }\href {\doibase
  10.1103/PhysRevC.94.034909} {\bibfield  {journal} {\bibinfo  {journal} {Phys.
  Rev.}\ }\textbf {\bibinfo {volume} {C94}},\ \bibinfo {pages} {034909}
  (\bibinfo {year} {2016})},\ \Eprint {http://arxiv.org/abs/1604.06212}
  {arXiv:1604.06212 [nucl-th]} \BibitemShut {NoStop}%
\bibitem [{\citenamefont {Westfall}(2015)}]{Westfall:2014fwa}%
  \BibitemOpen
  \bibfield  {author} {\bibinfo {author} {\bibfnamefont {G.~D.}\ \bibnamefont
  {Westfall}},\ }\href {\doibase 10.1103/PhysRevC.92.024902} {\bibfield
  {journal} {\bibinfo  {journal} {Phys. Rev.}\ }\textbf {\bibinfo {volume}
  {C92}},\ \bibinfo {pages} {024902} (\bibinfo {year} {2015})},\ \Eprint
  {http://arxiv.org/abs/1412.5988} {arXiv:1412.5988 [nucl-th]} \BibitemShut
  {NoStop}%
\bibitem [{\citenamefont {Feckov{\'a}}\ \emph {et~al.}(2015)\citenamefont
  {Feckov{\'a}}, \citenamefont {Steinheimer}, \citenamefont {Tom{\'a}{\v
  s}ik},\ and\ \citenamefont {Bleicher}}]{Feckova:2015qza}%
  \BibitemOpen
  \bibfield  {author} {\bibinfo {author} {\bibfnamefont {Z.}~\bibnamefont
  {Feckov{\'a}}}, \bibinfo {author} {\bibfnamefont {J.}~\bibnamefont
  {Steinheimer}}, \bibinfo {author} {\bibfnamefont {B.}~\bibnamefont
  {Tom{\'a}{\v s}ik}}, \ and\ \bibinfo {author} {\bibfnamefont
  {M.}~\bibnamefont {Bleicher}},\ }\href {\doibase 10.1103/PhysRevC.92.064908}
  {\bibfield  {journal} {\bibinfo  {journal} {Phys. Rev.}\ }\textbf {\bibinfo
  {volume} {C92}},\ \bibinfo {pages} {064908} (\bibinfo {year} {2015})},\
  \Eprint {http://arxiv.org/abs/1510.05519} {arXiv:1510.05519 [nucl-th]}
  \BibitemShut {NoStop}%
\bibitem [{\citenamefont {Bzdak}\ \emph {et~al.}(2016)\citenamefont {Bzdak},
  \citenamefont {Holzmann},\ and\ \citenamefont {Koch}}]{Bzdak:2016qdc}%
  \BibitemOpen
  \bibfield  {author} {\bibinfo {author} {\bibfnamefont {A.}~\bibnamefont
  {Bzdak}}, \bibinfo {author} {\bibfnamefont {R.}~\bibnamefont {Holzmann}}, \
  and\ \bibinfo {author} {\bibfnamefont {V.}~\bibnamefont {Koch}},\ }\href
  {\doibase 10.1103/PhysRevC.94.064907} {\bibfield  {journal} {\bibinfo
  {journal} {Phys. Rev.}\ }\textbf {\bibinfo {volume} {C94}},\ \bibinfo {pages}
  {064907} (\bibinfo {year} {2016})},\ \Eprint
  {http://arxiv.org/abs/1603.09057} {arXiv:1603.09057 [nucl-th]} \BibitemShut
  {NoStop}%
\bibitem [{\citenamefont {Braun-Munzinger}\ \emph {et~al.}(2017)\citenamefont
  {Braun-Munzinger}, \citenamefont {Rustamov},\ and\ \citenamefont
  {Stachel}}]{Braun-Munzinger:2016yjz}%
  \BibitemOpen
  \bibfield  {author} {\bibinfo {author} {\bibfnamefont {P.}~\bibnamefont
  {Braun-Munzinger}}, \bibinfo {author} {\bibfnamefont {A.}~\bibnamefont
  {Rustamov}}, \ and\ \bibinfo {author} {\bibfnamefont {J.}~\bibnamefont
  {Stachel}},\ }\href {\doibase 10.1016/j.nuclphysa.2017.01.011} {\bibfield
  {journal} {\bibinfo  {journal} {Nucl. Phys.}\ }\textbf {\bibinfo {volume}
  {A960}},\ \bibinfo {pages} {114} (\bibinfo {year} {2017})},\ \Eprint
  {http://arxiv.org/abs/1612.00702} {arXiv:1612.00702 [nucl-th]} \BibitemShut
  {NoStop}%
\bibitem [{\citenamefont {Bluhm}\ \emph {et~al.}(2017)\citenamefont {Bluhm},
  \citenamefont {Nahrgang}, \citenamefont {Bass},\ and\ \citenamefont
  {Schaefer}}]{Bluhm:2016byc}%
  \BibitemOpen
  \bibfield  {author} {\bibinfo {author} {\bibfnamefont {M.}~\bibnamefont
  {Bluhm}}, \bibinfo {author} {\bibfnamefont {M.}~\bibnamefont {Nahrgang}},
  \bibinfo {author} {\bibfnamefont {S.~A.}\ \bibnamefont {Bass}}, \ and\
  \bibinfo {author} {\bibfnamefont {T.}~\bibnamefont {Schaefer}},\ }\href
  {\doibase 10.1140/epjc/s10052-017-4771-3} {\bibfield  {journal} {\bibinfo
  {journal} {Eur. Phys. J.}\ }\textbf {\bibinfo {volume} {C77}},\ \bibinfo
  {pages} {210} (\bibinfo {year} {2017})},\ \Eprint
  {http://arxiv.org/abs/1612.03889} {arXiv:1612.03889 [nucl-th]} \BibitemShut
  {NoStop}%
\bibitem [{\citenamefont {Xu}(2017)}]{Xu:2016skm}%
  \BibitemOpen
  \bibfield  {author} {\bibinfo {author} {\bibfnamefont {H.-J.}\ \bibnamefont
  {Xu}},\ }\href {\doibase 10.1016/j.physletb.2016.12.015} {\bibfield
  {journal} {\bibinfo  {journal} {Phys. Lett.}\ }\textbf {\bibinfo {volume}
  {B765}},\ \bibinfo {pages} {188} (\bibinfo {year} {2017})},\ \Eprint
  {http://arxiv.org/abs/1612.06485} {arXiv:1612.06485 [nucl-th]} \BibitemShut
  {NoStop}%
\bibitem [{\citenamefont {Nonaka}\ \emph {et~al.}(2017)\citenamefont {Nonaka},
  \citenamefont {Kitazawa},\ and\ \citenamefont {Esumi}}]{Nonaka:2017kko}%
  \BibitemOpen
  \bibfield  {author} {\bibinfo {author} {\bibfnamefont {T.}~\bibnamefont
  {Nonaka}}, \bibinfo {author} {\bibfnamefont {M.}~\bibnamefont {Kitazawa}}, \
  and\ \bibinfo {author} {\bibfnamefont {S.}~\bibnamefont {Esumi}},\ }\href
  {\doibase 10.1103/PhysRevC.95.064912} {\bibfield  {journal} {\bibinfo
  {journal} {Phys. Rev.}\ }\textbf {\bibinfo {volume} {C95}},\ \bibinfo {pages}
  {064912} (\bibinfo {year} {2017})},\ \Eprint
  {http://arxiv.org/abs/1702.07106} {arXiv:1702.07106} \BibitemShut {NoStop}%
\bibitem [{\citenamefont {Garg}\ and\ \citenamefont
  {Mishra}(2017)}]{Garg:2017agr}%
  \BibitemOpen
  \bibfield  {author} {\bibinfo {author} {\bibfnamefont {P.}~\bibnamefont
  {Garg}}\ and\ \bibinfo {author} {\bibfnamefont {D.~K.}\ \bibnamefont
  {Mishra}},\ }\href@noop {} {\  (\bibinfo {year} {2017})},\ \Eprint
  {http://arxiv.org/abs/1705.01256} {arXiv:1705.01256 [nucl-th]} \BibitemShut
  {NoStop}%
\bibitem [{\citenamefont {Anticic}\ \emph {et~al.}(2010)\citenamefont {Anticic}
  \emph {et~al.}}]{Anticic:2009pe}%
  \BibitemOpen
  \bibfield  {author} {\bibinfo {author} {\bibfnamefont {T.}~\bibnamefont
  {Anticic}} \emph {et~al.} (\bibinfo {collaboration} {NA49}),\ }\href
  {\doibase 10.1103/PhysRevC.81.064907} {\bibfield  {journal} {\bibinfo
  {journal} {Phys. Rev.}\ }\textbf {\bibinfo {volume} {C81}},\ \bibinfo {pages}
  {064907} (\bibinfo {year} {2010})},\ \Eprint {http://arxiv.org/abs/0912.4198}
  {arXiv:0912.4198 [nucl-ex]} \BibitemShut {NoStop}%
\bibitem [{\citenamefont {Anticic}\ \emph {et~al.}(2015)\citenamefont {Anticic}
  \emph {et~al.}}]{Anticic:2012xb}%
  \BibitemOpen
  \bibfield  {author} {\bibinfo {author} {\bibfnamefont {T.}~\bibnamefont
  {Anticic}} \emph {et~al.} (\bibinfo {collaboration} {NA49}),\ }\href
  {\doibase 10.1140/epjc/s10052-015-3738-5} {\bibfield  {journal} {\bibinfo
  {journal} {Eur. Phys. J.}\ }\textbf {\bibinfo {volume} {C75}},\ \bibinfo
  {pages} {587} (\bibinfo {year} {2015})},\ \Eprint
  {http://arxiv.org/abs/1208.5292} {arXiv:1208.5292 [nucl-ex]} \BibitemShut
  {NoStop}%
\bibitem [{\citenamefont {Antoniou}\ \emph {et~al.}(2016)\citenamefont
  {Antoniou}, \citenamefont {Davis},\ and\ \citenamefont
  {Diakonos}}]{Antoniou:2015lwa}%
  \BibitemOpen
  \bibfield  {author} {\bibinfo {author} {\bibfnamefont {N.~G.}\ \bibnamefont
  {Antoniou}}, \bibinfo {author} {\bibfnamefont {N.}~\bibnamefont {Davis}}, \
  and\ \bibinfo {author} {\bibfnamefont {F.~K.}\ \bibnamefont {Diakonos}},\
  }\href {\doibase 10.1103/PhysRevC.93.014908} {\bibfield  {journal} {\bibinfo
  {journal} {Phys. Rev.}\ }\textbf {\bibinfo {volume} {C93}},\ \bibinfo {pages}
  {014908} (\bibinfo {year} {2016})},\ \Eprint
  {http://arxiv.org/abs/1510.03120} {arXiv:1510.03120 [hep-ph]} \BibitemShut
  {NoStop}%
\bibitem [{\citenamefont {Ling}\ and\ \citenamefont
  {Stephanov}(2016)}]{Ling:2015yau}%
  \BibitemOpen
  \bibfield  {author} {\bibinfo {author} {\bibfnamefont {B.}~\bibnamefont
  {Ling}}\ and\ \bibinfo {author} {\bibfnamefont {M.~A.}\ \bibnamefont
  {Stephanov}},\ }\href {\doibase 10.1103/PhysRevC.93.034915} {\bibfield
  {journal} {\bibinfo  {journal} {Phys. Rev.}\ }\textbf {\bibinfo {volume}
  {C93}},\ \bibinfo {pages} {034915} (\bibinfo {year} {2016})},\ \Eprint
  {http://arxiv.org/abs/1512.09125} {arXiv:1512.09125 [nucl-th]} \BibitemShut
  {NoStop}%
\bibitem [{\citenamefont {Bzdak}\ \emph
  {et~al.}(2017{\natexlab{a}})\citenamefont {Bzdak}, \citenamefont {Koch},\
  and\ \citenamefont {Strodthoff}}]{Bzdak:2016sxg}%
  \BibitemOpen
  \bibfield  {author} {\bibinfo {author} {\bibfnamefont {A.}~\bibnamefont
  {Bzdak}}, \bibinfo {author} {\bibfnamefont {V.}~\bibnamefont {Koch}}, \ and\
  \bibinfo {author} {\bibfnamefont {N.}~\bibnamefont {Strodthoff}},\ }\href
  {\doibase 10.1103/PhysRevC.95.054906} {\bibfield  {journal} {\bibinfo
  {journal} {Phys. Rev.}\ }\textbf {\bibinfo {volume} {C95}},\ \bibinfo {pages}
  {054906} (\bibinfo {year} {2017}{\natexlab{a}})},\ \Eprint
  {http://arxiv.org/abs/1607.07375} {arXiv:1607.07375 [nucl-th]} \BibitemShut
  {NoStop}%
\bibitem [{\citenamefont {Kitazawa}\ and\ \citenamefont
  {Luo}(2017)}]{Kitazawa:2017ljq}%
  \BibitemOpen
  \bibfield  {author} {\bibinfo {author} {\bibfnamefont {M.}~\bibnamefont
  {Kitazawa}}\ and\ \bibinfo {author} {\bibfnamefont {X.}~\bibnamefont {Luo}},\
  }\href@noop {} {\  (\bibinfo {year} {2017})},\ \Eprint
  {http://arxiv.org/abs/1704.04909} {arXiv:1704.04909 [nucl-th]} \BibitemShut
  {NoStop}%
\bibitem [{\citenamefont {Luo}(2015{\natexlab{b}})}]{Luo:2015ewa}%
  \BibitemOpen
  \bibfield  {author} {\bibinfo {author} {\bibfnamefont {X.}~\bibnamefont
  {Luo}} (\bibinfo {collaboration} {STAR}),\ }\bibfield  {booktitle} {\emph
  {\bibinfo {booktitle} {{Proceedings, 9th International Workshop on Critical
  Point and Onset of Deconfinement (CPOD 2014): Bielefeld, Germany, November
  17-21, 2014}}},\ }\href@noop {} {\bibfield  {journal} {\bibinfo  {journal}
  {PoS}\ }\textbf {\bibinfo {volume} {CPOD2014}},\ \bibinfo {pages} {019}
  (\bibinfo {year} {2015}{\natexlab{b}})},\ \Eprint
  {http://arxiv.org/abs/1503.02558} {arXiv:1503.02558 [nucl-ex]} \BibitemShut
  {NoStop}%
\bibitem [{\citenamefont {Luo}\ and\ \citenamefont {Xu}(2017)}]{Luo:2017faz}%
  \BibitemOpen
  \bibfield  {author} {\bibinfo {author} {\bibfnamefont {X.}~\bibnamefont
  {Luo}}\ and\ \bibinfo {author} {\bibfnamefont {N.}~\bibnamefont {Xu}},\
  }\href {\doibase 10.1007/s41365-017-0257-0} {\bibfield  {journal} {\bibinfo
  {journal} {Nucl. Sci. Tech.}\ }\textbf {\bibinfo {volume} {28}},\ \bibinfo
  {pages} {112} (\bibinfo {year} {2017})},\ \Eprint
  {http://arxiv.org/abs/1701.02105} {arXiv:1701.02105 [nucl-ex]} \BibitemShut
  {NoStop}%
\bibitem [{\citenamefont {Bzdak}\ \emph
  {et~al.}(2017{\natexlab{b}})\citenamefont {Bzdak}, \citenamefont {Koch},\
  and\ \citenamefont {Skokov}}]{Bzdak:2016jxo}%
  \BibitemOpen
  \bibfield  {author} {\bibinfo {author} {\bibfnamefont {A.}~\bibnamefont
  {Bzdak}}, \bibinfo {author} {\bibfnamefont {V.}~\bibnamefont {Koch}}, \ and\
  \bibinfo {author} {\bibfnamefont {V.}~\bibnamefont {Skokov}},\ }\href
  {\doibase 10.1140/epjc/s10052-017-4847-0} {\bibfield  {journal} {\bibinfo
  {journal} {Eur. Phys. J.}\ }\textbf {\bibinfo {volume} {C77}},\ \bibinfo
  {pages} {288} (\bibinfo {year} {2017}{\natexlab{b}})},\ \Eprint
  {http://arxiv.org/abs/1612.05128} {arXiv:1612.05128 [nucl-th]} \BibitemShut
  {NoStop}%
\bibitem [{\citenamefont {Jowzaee}()}]{star:a1}%
  \BibitemOpen
  \bibfield  {author} {\bibinfo {author} {\bibfnamefont {S.}~\bibnamefont
  {Jowzaee}} (\bibinfo {collaboration} {STAR}),\ }\href@noop {} {\bibinfo
  {journal} {Talk presented at the the XXVI international conference on
  ultrarelativistic heavy-ion collisions (Quark Matter 2017), Chicago, February
  2017}\ }\BibitemShut {NoStop}%
\bibitem [{\citenamefont {Lipiec}()}]{star:a2}%
  \BibitemOpen
\bibfield  {journal} {  }\bibfield  {author} {\bibinfo {author} {\bibfnamefont
  {A.}~\bibnamefont {Lipiec}} (\bibinfo {collaboration} {STAR}),\ }\href@noop
  {} {\bibinfo  {journal} {Talk presented at the XII Workshop on Particle
  Correlations and Femtoscopy, Amsterdam, June 2017}\ }\BibitemShut {NoStop}%
\bibitem [{\citenamefont {Llope}()}]{star:cpod}%
  \BibitemOpen
\bibfield  {journal} {  }\bibfield  {author} {\bibinfo {author} {\bibfnamefont
  {W.}~\bibnamefont {Llope}} (\bibinfo {collaboration} {STAR}),\ }\href@noop
  {} {\bibinfo  {journal} {Talk presented at the Conference on
      Critical Point and Onset of Deconfinement (CPOD 2017), Stony Brook, August 2017}\ }\BibitemShut {NoStop}%
\bibitem [{\citenamefont {Aihara}\ \emph {et~al.}(1986)\citenamefont {Aihara}
  \emph {et~al.}}]{Aihara:1986fy}%
  \BibitemOpen
\bibfield  {journal} {  }\bibfield  {author} {\bibinfo {author} {\bibfnamefont
  {H.}~\bibnamefont {Aihara}} \emph {et~al.} (\bibinfo {collaboration} {TPC/Two
  Gamma}),\ }\href {\doibase 10.1103/PhysRevLett.57.3140} {\bibfield  {journal}
  {\bibinfo  {journal} {Phys. Rev. Lett.}\ }\textbf {\bibinfo {volume} {57}},\
  \bibinfo {pages} {3140} (\bibinfo {year} {1986})}\BibitemShut {NoStop}%
\bibitem [{\citenamefont {Adam}\ \emph {et~al.}(2016)\citenamefont {Adam} \emph
  {et~al.}}]{Adam:2016iwf}%
  \BibitemOpen
  \bibfield  {author} {\bibinfo {author} {\bibfnamefont {J.}~\bibnamefont
  {Adam}} \emph {et~al.} (\bibinfo {collaboration} {ALICE}),\ }\href@noop {} {\
   (\bibinfo {year} {2016})},\ \Eprint {http://arxiv.org/abs/1612.08975}
  {arXiv:1612.08975 [nucl-ex]} \BibitemShut {NoStop}%
\bibitem [{\citenamefont {Bialas}\ \emph {et~al.}(1976)\citenamefont {Bialas},
  \citenamefont {Bleszynski},\ and\ \citenamefont {Czyz}}]{Bialas:1976ed}%
  \BibitemOpen
  \bibfield  {author} {\bibinfo {author} {\bibfnamefont {A.}~\bibnamefont
  {Bialas}}, \bibinfo {author} {\bibfnamefont {M.}~\bibnamefont {Bleszynski}},
  \ and\ \bibinfo {author} {\bibfnamefont {W.}~\bibnamefont {Czyz}},\ }\href
  {\doibase 10.1016/0550-3213(76)90329-1} {\bibfield  {journal} {\bibinfo
  {journal} {Nucl. Phys.}\ }\textbf {\bibinfo {volume} {B111}},\ \bibinfo
  {pages} {461} (\bibinfo {year} {1976})}\BibitemShut {NoStop}%
\bibitem [{\citenamefont {Bzdak}\ and\ \citenamefont
  {Bozek}(2016)}]{Bzdak:2015dja}%
  \BibitemOpen
  \bibfield  {author} {\bibinfo {author} {\bibfnamefont {A.}~\bibnamefont
  {Bzdak}}\ and\ \bibinfo {author} {\bibfnamefont {P.}~\bibnamefont {Bozek}},\
  }\href {\doibase 10.1103/PhysRevC.93.024903} {\bibfield  {journal} {\bibinfo
  {journal} {Phys. Rev.}\ }\textbf {\bibinfo {volume} {C93}},\ \bibinfo {pages}
  {024903} (\bibinfo {year} {2016})},\ \Eprint
  {http://arxiv.org/abs/1509.02967} {arXiv:1509.02967 [hep-ph]} \BibitemShut
  {NoStop}%
\bibitem [{\citenamefont {Adamczyk}\ \emph {et~al.}(2017)\citenamefont
  {Adamczyk} \emph {et~al.}}]{Adamczyk:2017iwn}%
  \BibitemOpen
  \bibfield  {author} {\bibinfo {author} {\bibfnamefont {L.}~\bibnamefont
  {Adamczyk}} \emph {et~al.} (\bibinfo {collaboration} {STAR}),\ }\href@noop {}
  {\  (\bibinfo {year} {2017})},\ \Eprint {http://arxiv.org/abs/1701.07065}
  {arXiv:1701.07065 [nucl-ex]} \BibitemShut {NoStop}%
\bibitem [{\citenamefont {Anticic}\ \emph {et~al.}(2011)\citenamefont {Anticic}
  \emph {et~al.}}]{Anticic:2010mp}%
  \BibitemOpen
  \bibfield  {author} {\bibinfo {author} {\bibfnamefont {T.}~\bibnamefont
  {Anticic}} \emph {et~al.} (\bibinfo {collaboration} {NA49}),\ }\href
  {\doibase 10.1103/PhysRevC.83.014901} {\bibfield  {journal} {\bibinfo
  {journal} {Phys. Rev.}\ }\textbf {\bibinfo {volume} {C83}},\ \bibinfo {pages}
  {014901} (\bibinfo {year} {2011})},\ \Eprint {http://arxiv.org/abs/1009.1747}
  {arXiv:1009.1747 [nucl-ex]} \BibitemShut {NoStop}%
\end{thebibliography}
\end{document}